\documentclass[11pt]{article}
\usepackage{amssymb,amsmath,amsfonts,graphicx,hyperref}

\input{tcilatex}
\begin{document}

\title{Sine--Gordon Solitons, Kinks and Breathers as Physical Models of Nonlinear Excitations in Living Cellular Structures}
\author{Vladimir G. Ivancevic\thanks{Vladimir.Ivancevic@dsto.defence.gov.au} ~and Tijana T. Ivancevic\thanks{Tijana.Ivancevic@alumni.adelaide.edu.au}}\date{}
\maketitle

\vspace{2cm}

\hyphenation{equations}
\begin{abstract}
\noindent Nonlinear space-time dynamics, defined in terms of celebrated `solitonic' equations, brings indispensable tools for understanding, prediction and control of complex behaviors in both physical and life sciences. In this paper, we review sine--Gordon solitons, kinks and breathers as models of nonlinear excitations in complex systems in physics and in living cellular structures, both intra--cellular (DNA, protein folding and microtubules) and inter--cellular (neural impulses and muscular contractions).\\~\\

\end{abstract}

\newpage
\tableofcontents

\newpage
\section{Introduction}

\begin{figure}[h]
\centerline{\includegraphics[width=9.5cm]{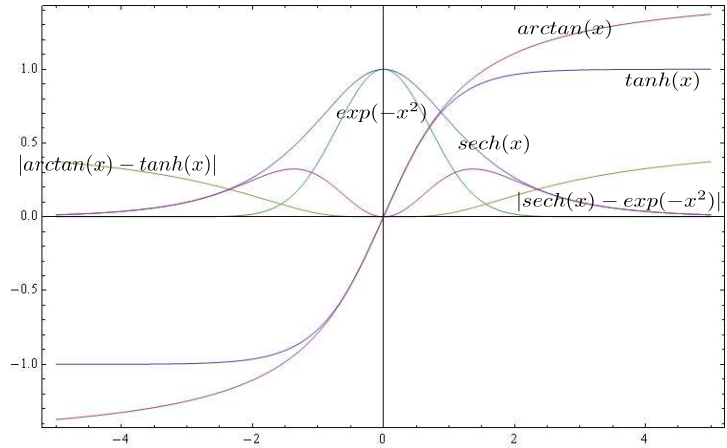}}
\caption{Basic static examples of kinks: $\func{tanh}(x),~\func{arctan}(x)$
and bell--shaped solitons: $\func{sech}(x),~\func{exp}(-x^2)$, ~together with their
(absolute) differences; plotted in $Mathematica^{TM}$.}
\label{solitons}
\end{figure}

In spatiotemporal dynamics of complex nonlinear systems (see \cite{coment,IvTurbul,IvRicci,Complexity}), Sine--Gordon equation (SGE) is, together with Korteweg--deVries (KdV) and
nonlinear Schr\"{o}dinger (NLS) equations, one of the celebrated
nonlinear-yet-integrable partial differential equations (PDEs),\footnote{
For a soft SGE--intro, see popular web-sites: \cite%
{Wiki,Wolfram,Encyclopedia}. Also, both KdV and NLS equations are mentioned in
subsection \ref{Poisson} below as solitary models for muscular contractions on
Poisson manifolds.} with a variety of traveling solitary waves as solutions\footnote{%
A solitary wave is a traveling wave (with velocity $v$) of the form: $\phi
(x,t)=f(x-vt)$, for a smooth function $f$ that decays rapidly at infinity;
e.g., a nonlinear wave equation: ~$\phi _{tt}-\phi _{xx}=\phi (2\phi ^{2}-1)$%
~ has a family of solitary--wave solutions: $\phi (x,t)=\mathrm{sech}(x%
\mathrm{cosh}\mu +t\mathrm{sinh}\mu ),$ parameterized by $\mu \in \mathbb{R}$
(see \cite{Terng}).} (see Figures \ref{solitons} and \ref{solitPendula}, as
well as the following basic references: \cite%
{Bullough,Ablowitz81,Rajaraman,Novikov,Newell85,Polyanin,Faddeev}). In complex
physical systems, SGE solitons, kinks and breathers appear in various situations,
including propagation of magnetic flux (fluxons) in long Josephson junctions
\cite{Kivshar,Braun98}, dislocations in crystals \cite{FrenKont,Braun04},
nonlinear spin waves in superfluids \cite{Kivshar}, and waves in ferromagnetic and
anti-ferromagnetic materials \cite{ZharnitskyPRB,ZharnitskyPRE} -- to
mention just a few application areas.

In this paper, we review physical theory of sine--Gordon solitons, kinks and breathers, as well as their essential dynamics of nonlinear excitations in living cellular structures, both intra--cellular (DNA, protein folding and microtubules) and inter--cellular (neural impulses and muscular contractions).

We show that sine--Gordon traveling waves can give us new insights even in such long--time established and Nobel--Prize winning living systems as the Watson--Crick double helix DNA model and the Hodgkin--Huxley neural conduction model.

\section{Physical theory of sine--Gordon solitons, kinks and breathers}

In this section, we give the basic theory of the sine--Gordon equation (and the variety of its traveling--wave solutions), as spatiotemporal models of nonlinear excitations in complex physical systems.

\subsection{Sine--Gordon equation (SGE)}

SGE is a real-valued, hyperbolic, nonlinear wave equation defined on $%
\mathbb{R}^{1,1}$, which appears in two equivalent forms (using standard
indicial notation for partial derivatives: $\phi _{zz}=\partial _{z}^{2}\phi
=\partial ^{2}\phi /\partial x^{z}$):

\begin{itemize}
\item In the (1+1) space-time $(x,t)-$coordinates, the SGE reads:
\begin{equation}
\phi _{tt}=\phi _{xx}-\sin \phi ,\qquad \text{\textrm{or}\qquad }\phi
_{tt}(x,t)=\phi _{xx}(x,t)-\sin \phi (x,t),  \label{SGE1}
\end{equation}%
which shows that it is a nonlinear extension of the standard linear wave
equation: $\phi _{tt}=\phi _{xx}.$ The solutions $\phi (x,t)$\ of (\ref{SGE1}%
) determine the internal Riemannian geometry of surfaces of constant
negative scalar curvature $R=-2,$ given by the line-element:%
\begin{equation*}
ds^{2}=\sin ^{2}\left( \frac{\phi }{2}\right) dt^{2}+\cos ^{2}\left( \frac{%
\phi }{2}\right) dx^{2},
\end{equation*}%
where the angle $\phi $ describes the embedding of the surface into
Euclidean space $\mathbb{R}^{3}$ (see \cite{Bullough}). A basic solution of
the SGE (\ref{SGE1}) is:%
\begin{equation}
\phi (x,t)=4\arctan \left[ \exp \left( \pm \frac{x-vt}{\sqrt{1-v^{2}}}%
\right) \right] ,  \label{fund}
\end{equation}%
describing a \emph{soliton} moving with velocity $0\leq v<1$ and changing
the phase from 0 to 2$\pi $ (\emph{kink,} the case of $+$ sign) or from 2$%
\pi $ to 0 (\emph{anti-kink,} the case of $-$ sign).\footnote{%
Each traveling soliton solution of the SGE has the corresponding surface in $%
\mathbb{R}^3$ (see \cite{Terng}).}

\item In the (1+1) light-cone $(u,v)-$coordinates, defined by: $%
u=(x+t)/2,\,v=(x+t)/2$, in which the line-element (depending on the angle $%
\phi $\ between two asymptotic lines: $u=\mathrm{const},v=\mathrm{const}$)
is given by: \
\begin{equation*}
ds^{2}=du^{2}+2\cos \phi \,du\,dv+dv^{2},
\end{equation*}%
the SGE describes a family of pseudo-spherical surfaces with constant
Gaussian curvature $K=-1,$ and reads:\footnote{%
SGE (\ref{SGE2}) is the single Codazzi--Mainardi compatibility equation
between the first $(I_{\mathrm{G}})$ and second $(II_{\mathrm{C}})$
fundamental forms of a surface, defined by the Gauss and Codazzi equations,
respectively:
\par
\begin{equation*}
I_{\mathrm{G}}=du^{2}+2\cos \phi \,du\,dv+dv^{2},\qquad II_{\mathrm{C}%
}=2\sin \phi \,du\,dv.
\end{equation*}%
}
\begin{equation}
\phi _{uv}=\sin \phi \,,\qquad \text{\textrm{or}\qquad }\phi _{uv}(u,v)=\sin
\phi (u,v).  \label{SGE2}
\end{equation}
\end{itemize}

\begin{figure}[h]
\centerline{\includegraphics[width=13cm]{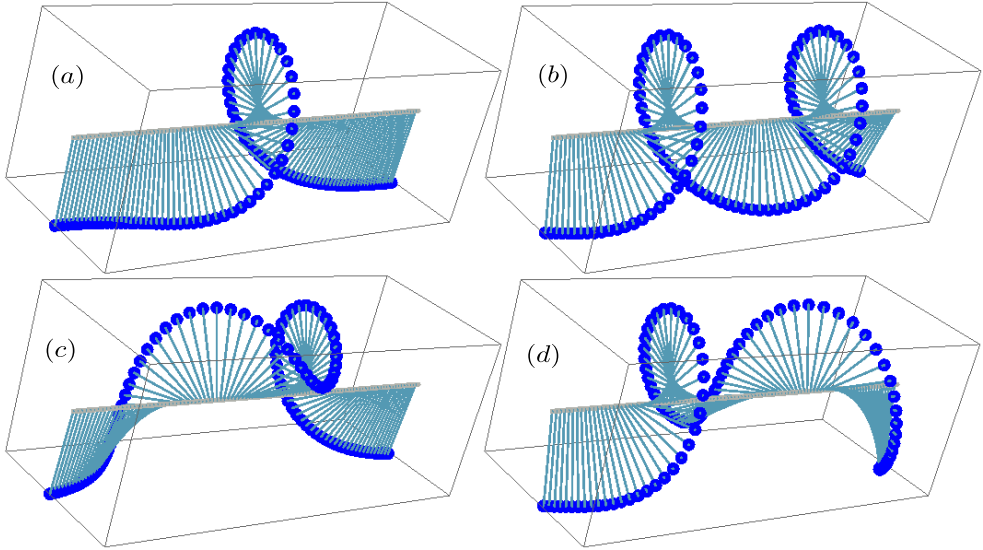}}
\caption{Basic solitary SGE--solutions, simulated in $Mathematica$ as systems of
spring-coupled torsional pendula: (a) single soliton: $\protect\phi %
(x,t)=4\arctan \left( \exp \frac{x-vt}{1-v^{2}}\right) ; $ (b)
soliton--soliton collision: $\protect\phi (x,t)=4\arctan \left( \frac{v\sinh
\frac{x}{1-v^{2}}}{\cosh \frac{vt}{1-v^{2}}}\right) ; $ (c)
soliton--antisoliton collision: $\protect\phi (x,t)=4\arctan \left( \frac{%
\sinh \frac{vt}{1-v^{2}}}{v\cosh \frac{x}{1-v^{2}}}\right) ; $ and (d)
single breather: $\protect\phi (x,t)=4\arctan \left( \frac{\sin \frac{vt}{%
1-v^{2}}}{v\cosh \frac{x}{1-v^{2}}}\right)$ (modified and adapted from \cite{Meszena}).}
\label{solitPendula}
\end{figure}

\begin{figure}[h]
\centerline{\includegraphics[width=12cm]{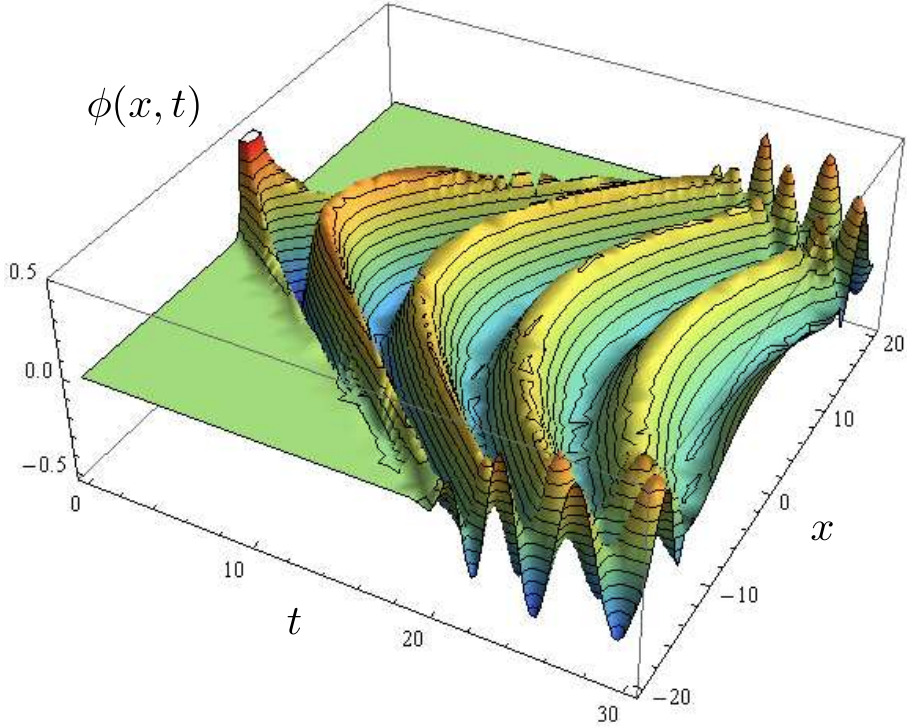}}
\caption{Numerical solution of the SGE (\protect\ref{SGE1}) in $Mathematica$, using numerical ODE/PDE integrator NDSolve, with the
following data (including the Gaussian initial state, zero initial velocity
and symmetric boundary condition): $x\in \lbrack -20,20],~~ t\in [0,30],~~
\protect\phi (x,0)=\exp({-x^2}),~~ \protect\phi _{t}(x,0)=0,~~ \protect\phi %
(-20,t)=\protect\phi (20,t)$. The waves oscillate around the zero plane and
increase their width with time. Both near-periodicity and nonlinearity of
the time evolution are apparent.}
\label{sgMma}
\end{figure}

A typical, spatially-symmetric, boundary-value problem for (\ref{SGE1}) is
defined by:%
\begin{eqnarray*}
x &\in &[-L,L]\subset \mathbb{R},\qquad \left( t\in \mathbb{R}^{+}\right) ,%
\hspace{1.7cm} \\
\phi (x,0) &=&f(x),\qquad \phi _{t}(x,0)=0,\qquad \phi (-L,t)=\phi (L,t),
\end{eqnarray*}%
where $f(x)\in \mathbb{R}$ is an axially-symmetric function (e.g., Gaussian
or sech, see Figure \ref{sgMma}).

\emph{B\"{a}cklund transformations} (BT) for the SGE (\ref{SGE1}) were
devised in 1880s in Riemannian geometry of surfaces and are attributed to
Bianchi and B\"{a}cklund\footnote{%
In 1883, A. B\"{a}cklund showed that if $L:M\to M^{\prime }$ is a
pseudo-spherical line congruence between two surfaces $M,M^{\prime }$, then
both $M$ and $M^{\prime }$ are pseudo-spherical and $L$ maps asymptotic
lines on $M$ to asymptotic lines on $M^{\prime }$. Analytically, this is
equivalent to the statement that if $\phi$ is a solution of the SGE (\ref%
{SGE1}), then so are also the solutions of the ODE system (\ref{bt}).} (see,
e.g. \cite{Darboux}). They have the form:
\begin{equation}
\frac{1}{2}(\phi +\varphi )_{\xi }=\alpha \sin \frac{\phi -\varphi }{2}%
,\qquad \frac{1}{2}(\phi -\varphi )_{\eta }=\frac{1}{\alpha }\sin \frac{\phi
+\varphi }{2},  \label{bt}
\end{equation}%
where both $\phi $ and $\varphi $ are solutions of the SGE (\ref{SGE1}), and
can be viewed as a transformation of the SGE into itself. BT (\ref{bt})
allows one to find a 2-parameter family of solutions, given a particular
solution $\phi _{0}$ of (\ref{SGE1}). For example, consider the trivial
solution $\phi =0$ that, substituted into (\ref{bt}), gives:%
\begin{equation*}
\varphi _{\xi }=-2\alpha \sin \frac{\varphi }{2},\qquad \varphi _{\eta }=-%
\frac{2}{\alpha }\sin \frac{\varphi }{2},
\end{equation*}%
which, by integration, gives:
\begin{equation*}
2\alpha \xi =-2\ln (\tan \frac{\varphi }{4})+p(\eta ),\qquad \frac{2}{\alpha
}\eta =-2\ln (\tan \frac{\varphi }{4})+p(\xi ),
\end{equation*}%
from which the following new solution is generated:%
\begin{equation*}
\varphi =4\arctan \left[ \exp (-\alpha \xi -\frac{1}{\alpha }\eta +\mathrm{%
const})\right] .
\end{equation*}

The sine--forcing term in the SGE can be viewed as a \emph{nonlinear
deformation:} $\phi \rightarrow \sin \phi ,$ of the linear forcing term in
the Klein--Gordon equation (KGE, a vacuum linearization of the SGE), which
is commonly used for describing scalar fields (quantum) field theory:
\begin{equation}
\phi _{tt}=\phi _{xx}-\phi ,  \label{KG}
\end{equation}%
This, in turn, implies that (as a field equation) SGE can be derived as an
Euler--Lagrangian equation from the Lagrangian density:
\begin{equation}
\mathcal{L}_{\text{SG}}(\phi )=\frac{1}{2}(\phi _{t}^{2}-\phi
_{x}^{2})-1+\cos \phi .  \label{LSG}
\end{equation}%
It could be expected that $\mathcal{L}_{\text{SG}}(\phi )$ is a
`deformation' of the KG Lagrangian:%
\begin{equation}
\mathcal{L}_{\text{KG}}(\phi )=\frac{1}{2}(\phi _{t}^{2}-\phi _{x}^{2})-%
\frac{\phi ^{2}}{2}.  \label{LKG}
\end{equation}%
That can be demonstrated by the Taylor--series expansion of the cosine term:
\begin{equation*}
\cos \phi =\sum_{n=0}^{\infty }\frac{(-\phi ^{2})^{n}}{(2n)!},
\end{equation*}%
so that we have the following relationship between the two Lagrangians:
\begin{equation*}
\mathcal{L}_{\text{SG}}(\phi )=\mathcal{L}_{\text{KG}}(\phi
)+\sum_{n=2}^{\infty }\frac{(-\phi ^{2})^{n}}{(2n)!}.
\end{equation*}%
\begin{figure}[h]
\centerline{\includegraphics[width=11.5cm]{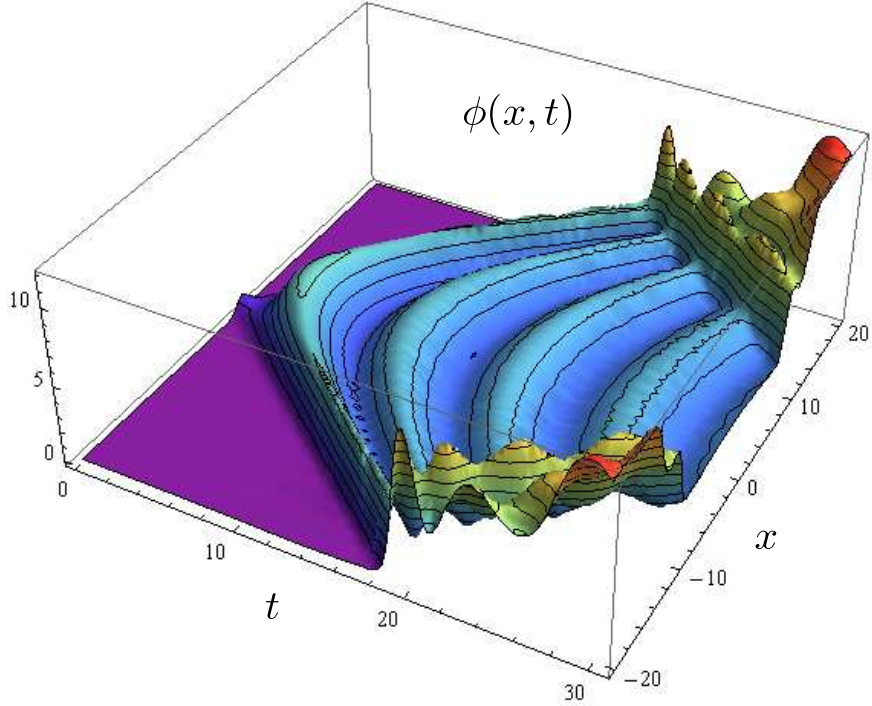}}
\caption{Numerical solution of the SGE (\protect\ref{SGE+}) in $Mathematica$%
, with the following data (including the Gaussian initial state, zero
initial velocity and symmetric boundary condition): $x\in \lbrack -20,20],~~
t\in [0,30],~~ \protect\phi (x,0)=\exp({-x^2}),~~ \protect\phi_{t}(x,0)=0,~~
\protect\phi (-20,t)=\protect\phi (20,t)$. Under the same boundary
conditions, the SGE with the plus sine gives about 20 times higher amplitude
waves, which are all above the zero plane and decrease their width with
time. Again, both near-periodicity and nonlinearity of the time evolution
are apparent.}
\label{sgMma2}
\end{figure}

The corresponding Hamiltonian densities, of kinetic plus potential energy
type, are given in terms of canonically--conjugated coordinate and momentum
fields by:
\begin{eqnarray*}
\mathcal{H}_{\text{SG}}(\phi ,\pi ) &=&\pi \phi _{t}-\mathcal{L}_{\text{SG}%
}(\phi )=\frac{1}{2}(\pi ^{2}+\phi _{x}^{2})+1-\cos \phi , \\
\mathcal{H}_{\text{KG}}(\phi ,\pi ) &=&\pi \phi _{t}-\mathcal{L}_{\text{KG}%
}(\phi )=\frac{1}{2}(\pi ^{2}+\phi _{x}^{2})+\phi ^{2}.
\end{eqnarray*}%
Both SGE and KGE are infinite--dimensional Hamiltonian systems \cite{Temam},
with Poisson brackets given by:
\begin{equation}
\left\{ F,G\right\} =\int_{-\infty }^{\infty }\left[ \frac{\delta F}{\delta
\phi (x)}\frac{\delta G}{\delta \pi (x)}-\frac{\delta F}{\delta \pi (x)}%
\frac{\delta G}{\delta \phi (x)}\right] dx,  \label{PB}
\end{equation}%
so that both (\ref{SGE1}) and (\ref{KG}) follow from Hamilton's equations
with Hamiltonian $H$ and symplectic form $\omega $:\footnote{%
For the Poisson--manifold generalization, see section \ref{Poisson} below.}
\begin{eqnarray}
\phi _{t} &=&\{H,\phi \},\qquad \pi _{t}=\{H,\pi \},\qquad  \notag \\
\text{\textrm{with} \ \ \ }H &=&\int_{-\infty }^{\infty }\mathcal{H}(\phi
,\pi )\,dx,\qquad \omega =\int_{-\infty }^{\infty }d\pi \wedge d\phi \,\,dx.
\label{H}
\end{eqnarray}%
The Hamiltonian (\ref{H}) is conserved by the flow of both SGE (\ref{SGE1})
and KGE (\ref{KG}), with an infinite number of commuting constants of motion
(common level sets of these constants of motion are generically
infinite-dimensional tori of maximal dimension). Both SGE and KGE admit
their own infinite families of conserved functionals in involution with
respect to their Poisson bracket (\ref{PB}). This fact allows them both to
be solved with the \emph{inverse scattering transform} (see \cite{Faddeev}).

\subsection{Momentum and energy of SGE--solitons}

SGE is Lorentz--covariant (i.e., invariant with respect to
special--relativistic Lorentz transformations; each SGE--soliton behaves as
a relativistic object and contracts when $v\rightarrow c\equiv $ the speed
of light), and for this fact it has been used in (quantum) field theory.%
\footnote{%
The SGE, in both forms (\ref{SGE1}) and (\ref{SGE2}) has the following
symmetries:
\begin{eqnarray*}
&&%
\begin{array}{cccc}
t\rightarrow t+t_{0}, & x\rightarrow x, & \phi \rightarrow \phi & (\mathrm{%
shift\ in\ }t), \\
t\rightarrow t, & x\rightarrow x+x_{0}, & \phi \rightarrow \phi & (\mathrm{%
shift\ in\ }x), \\
t\rightarrow t, & x\rightarrow x, & \phi \rightarrow \phi +2\pi n & (\mathrm{%
discrete\ shifts\ in\ }\phi ), \\
t\rightarrow -t, & x\rightarrow x, & \phi \rightarrow \phi & (\mathrm{%
reflection\ in\ }t), \\
t\rightarrow t, & x\rightarrow -x, & \phi \rightarrow \phi & (\mathrm{%
reflection\ in\ }x), \\
t\rightarrow t, & x\rightarrow x, & \phi \rightarrow -\phi & (\mathrm{%
reflection\ in\ }\phi ),%
\end{array}
\\
&&%
\begin{array}{cccc}
t\rightarrow \frac{t-vx}{\sqrt{1-v^{2}}}, & x\rightarrow \frac{x-vx}{\sqrt{%
1-v^{2}}}, & \phi \rightarrow \phi & (\text{\textrm{Lorentz transformations
with velocity }}v),%
\end{array}%
\end{eqnarray*}%
where e.g. reflection in $\phi $ means: if $\phi $\ is a solution then so is
$-\phi $, etc.} In Minkowski (1+1$\mathbb{)}$ space-time coordinates ($%
x^{\mu }\in \mathbb{R}^{1,1},\,x^{0}=t,\,x^{1}=x)$ with metric tensor $\eta
_{\mu \nu }$ ($\mu ,\nu =0,1;\eta _{11}=-\eta _{22}=1,\eta _{11}=\eta
_{11}=0 $), the SG--Lagrangian density has the following `massive form' of
kinetic minus potential energy, with mass $m$ and coupling constant $\lambda
$ (see \cite{Rajaraman}):
\begin{equation*}
\mathcal{L}_{\text{SG}}^{\mathrm{Mink}}(\phi )=\frac{1}{2}(\phi
_{t}^{2}-\phi _{x}^{2})-\frac{m^{4}}{\lambda }\left[ 1-\cos \left( \frac{%
\sqrt{\lambda }}{m}\phi \right) \right] \,,
\end{equation*}%
which reduces to the dimensionless form (\ref{LSG}) by re-scaling the fields
and coordinates:
\begin{equation}
\frac{\sqrt{\lambda }}{m}\phi \rightarrow \phi ,\qquad mx^{\mu }\rightarrow
x^{\mu }.  \label{resc}
\end{equation}

The SG--Lagrangian density $\mathcal{L}_{\text{SG}}^{\mathrm{Mink}}(\phi
)\equiv \frac{m^{4}}{\lambda }\mathcal{L}_{\text{SG}}(\phi )$ obeys the
conservation law and admits topological\footnote{%
The word topological means that it is not sensitive to local degrees of
freedom.} \emph{Noether current} [with respect to (\ref{resc})]:%
\begin{equation*}
j^{\mu }=\frac{1}{2\pi }\varepsilon ^{\mu \nu }\partial _{\nu }\phi \quad
\text{with zero-divergence: \ \ }\partial _{\mu }j^{\mu }=0,
\end{equation*}%
where $\varepsilon ^{\mu \nu }$ is the $\mathbb{R}^{1,1}-$Levi--Civita
tensor. The corresponding topological \emph{Noether charge} is given by:
\begin{eqnarray*}
Q &=&\int \left\vert \partial _{t}j^{0}(x,t)\right\vert dx=\frac{1}{2\pi }%
\left\vert \phi (+\infty ,t)-\phi (-\infty ,t)\right\vert , \\
\text{\textrm{with} \ }Q_{t} &=&\frac{1}{2\pi }\left\vert \phi _{t}(-\infty
,t)-\phi _{t}(+\infty ,t)\right\vert =0.
\end{eqnarray*}

The most important physical quality of SGE is its energy--momentum (EM)
tensor $T_{\mu \nu }$, which is the Noether current corresponding to
spacetime--translation symmetry: $x^{\mu }\rightarrow x^{\mu }+\xi ^{\mu };$
this conserved quantity is derived from the Lagrangian (\ref{LSG}) as:%
\begin{equation*}
T_{\mu \nu }=\partial _{\mu }\phi \partial _{\nu }\phi -\eta _{\mu \nu }%
\mathcal{L}_{\text{SG}}(\phi ).
\end{equation*}%
$T_{\mu \nu }$ has the following components \cite{Rajaraman,Guilarte}:%
\begin{eqnarray*}
T_{00} &=&\frac{1}{2}(\phi _{t}^{2}+\phi _{x}^{2})+1-\cos \phi ,\qquad
T_{10}=\phi _{xt}\,=T_{01}. \\
T_{11} &=&\frac{1}{2}(\phi _{t}^{2}+\phi _{x}^{2})-1+\cos \phi ,
\end{eqnarray*}%
EM's contravariant form $T^{\mu \nu }$ has the following components:%
\footnote{%
The contravariant EM $T^{\mu \nu }$is obtained by raising the indices of $%
T_{\mu \nu }$ using the inverse metric tensor $\eta ^{\mu \nu }=1/(\eta
)_{\mu \nu }.$}%
\begin{equation*}
T^{00}=T_{00},\qquad T^{11}=T_{11},\qquad T^{10}=-T_{01}.
\end{equation*}%
EM's conserved quantities are: \emph{momentum} $P=\int T^{10}dx$, which is
the Noether charge with respect to space--translation symmetry, and \emph{%
energy} $E=\int T^{00}dx$, which is the Noether charge with respect to
time--translation symmetry. Energy and momentum follow from EM's zero
divergence:%
\begin{equation*}
\partial _{\mu }T^{\mu \nu }=0\Longrightarrow \left\{
\begin{array}{c}
\partial _{t}T^{00}-\partial _{x}T^{10}=0 \\
\partial _{t}T^{01}-\partial _{x}T^{11}=0%
\end{array}%
\right. \Longrightarrow \left\{
\begin{array}{c}
\partial _{t}E=\partial _{t}\int T^{00}dx=0 \\
\partial _{t}P=\partial _{t}\int T^{10}dx=0%
\end{array}%
\right. .
\end{equation*}

\subsection{SGE solutions and integrability}

\subsubsection{SGE solitons, kinks and breathers}
\label{kinks}

The first \emph{1-soliton} solution of the SGE (\ref{SGE1}) was given by
\cite{Ablowitz73,AblowitzBk} in the form:
\begin{equation*}
\phi (x,t)=4\arctan \left[ \frac{\sqrt{1-\omega ^{2}}\cos (\omega t)}{\omega
\cosh (x\sqrt{1-\omega ^{2}})}\right] ,
\end{equation*}%
which, for $\omega <1$, is periodic in time $t$ and decays exponentially
when moving away from $x=0$.

There is a well-known traveling solitary wave solution with velocity $v$
(see \cite{Tabor}), given by the following generalization of (\ref{fund}):%
\begin{equation}
\phi (x,t)=4\arctan \left[ \exp \frac{\pm 2(z-z_{0})}{\sqrt{1-v^{2}}\;}%
\right] ,\qquad \text{\textrm{with} \ \ }\left( z=\mu (x+vt)\right) ,
\label{tabor}
\end{equation}%
\ and the center at $z_{0}.$ In (\ref{tabor}), the case $+2$ describes kink,
while the case $-2$ corresponds to antikink.

The \emph{stationary kink} with the center at $x_{0}$ is defined by:
\begin{equation*}
\phi (x)=2\arctan \left[ \exp (x-x_{0})\right] ,
\end{equation*}%
(in which the position of the center $x_{0}$ can be varied continuously: $%
-\infty <x_{0}<\infty $) and represents the solution of the first-order ODE:
\ $\phi _{x}(x)=\sin \phi (x).$

Regarding solutions of the slightly more general, three-parameter SGE:
\begin{equation}
\phi _{tt}=a\phi _{xx}+b\sin (\lambda \phi ),  \label{SGen}
\end{equation}%
the following cases were established in the literature (see \cite{Polyanin}
and references therein):

\begin{enumerate}
\item If a function $w=\phi (x,t)$ is a solution of (\ref{SGen}), then so
are also the following functions:
\begin{eqnarray*}
w_{1} &=&\frac{2\pi n}{b}\pm \phi (C_{1}\pm x,C_{2}\pm t)\qquad \text{%
\textrm{for} \ \ }\left( n=0,\pm 1,\pm 2,...\right) , \\
w_{2} &=&\pm \phi \left( x\cosh C_{3}+t\sqrt{a}\sinh C_{3},\,\,x\frac{\sinh
C_{3}}{\sqrt{a}}+t\cosh C_{3}\right) ,
\end{eqnarray*}%
where $C_{1}$, $C_{2}$, and $C_{3}$ are arbitrary constants.

\item Traveling-wave solutions:
\begin{eqnarray}
\phi (x,t) &=&\frac{4}{\lambda }\arctan \left[ \exp \left( \pm \frac{%
b\lambda (C_{1}x+C_{2}t+C_{3})}{\sqrt{b\lambda (C_{2}^{2}-aC_{1}^{2})}}%
\right) \right] \quad  \label{atan1} \\
\text{\textrm{if} \ }b\lambda (C_{2}^{2}-aC_{1}^{2}) &>&0,  \notag \\
\phi (x,t) &=&-\frac{\pi }{\lambda }+\frac{4}{\lambda }\arctan \left[ \exp
\left( \pm \frac{b\lambda (C_{1}x+C_{2}t+C_{3})}{\sqrt{b\lambda
(aC_{1}^{2}-C_{2}^{2})}}\right) \right] \quad  \notag \\
\text{\textrm{if} \ }b\lambda (C_{2}^{2}-aC_{1}^{2}) &<&0,  \notag
\end{eqnarray}%
where the first expression (for $b\lambda (C_{2}^{2}-aC_{1}^{2})>0$)
represents another 1-soliton solution, which is kink in case of $\ \exp
\left( \frac{b\lambda (C_{1}x+C_{2}t+C_{3})}{\sqrt{b\lambda
(C_{2}^{2}-aC_{1}^{2})}}\right) $ \ and antikink in case of $\ \exp \left( -%
\frac{b\lambda (C_{1}x+C_{2}t+C_{3})}{\sqrt{b\lambda (C_{2}^{2}-aC_{1}^{2})}}%
\right) .$ In case of the standard SGE (\ref{SGE1}), this kink--antikink
expression specializes to the Lorentz-invariant solution similar to (\ref%
{tabor}):%
\begin{equation}
\phi ^{\mathrm{K}}(x,t)=4\arctan \left[ \exp \left( \frac{\pm (x-x_{c})-vt}{%
\sqrt{1-v^{2}}}\right) \right] ,  \label{kink1}
\end{equation}%
where the velocity $v$ ($0<v<1$) and the soliton-center $x_{c}$ are
real-valued constants. The kink solution has the following physical (EM)
characteristics:

(i) Energy:
\begin{equation*}
E[\phi ^{\mathrm{K}}(x,t)]=\int T^{00}dx=\frac{8}{\sqrt{1-v^{2}}};
\end{equation*}

(ii) Momentum:
\begin{equation*}
P[\phi ^{\mathrm{K}}(x,t)]=\int T^{10}dx=-\frac{8v}{\sqrt{1-v^{2}}}.
\end{equation*}

\item Functional separable solution:
\begin{equation*}
w(x,t)=\frac{4}{\lambda }\arctan \left[ f(x)g(t)\right] ,
\end{equation*}%
where the functions $f=f(x)$ and $g=g(t)$ are determined by the first-order
autonomous separable ODEs:
\begin{equation*}
f_{x}^{2}=Af^{4}+Bf^{2}+C,\qquad g_{t}^{2}=-aCg^{4}+(aB+b\lambda )g^{2}-aA,
\end{equation*}%
where $A$, $B$, and $C$ are arbitrary constants. In particular, for $%
A=0,B=k^{2}>0$, and $C>0$, we have the 2\emph{-soliton} solution of \cite%
{Perring62}: \ \
\begin{equation*}
w(x,t)=\frac{4}{\lambda }\arctan \left[ \frac{\eta \sin (kx+A_{1})}{k\sqrt{a}%
\cosh (\eta t+B_{1})}\right] ,\qquad \text{\textrm{with} \ \ }\left( \eta
^{2}=ak^{2}+b\lambda >0\right) ,
\end{equation*}%
where $k$, $A_{1}$, and $B_{1}$ are arbitrary constants.
\end{enumerate}

The only stable traveling wave SGE-solutions for a scalar field $\phi $ are $%
2\pi $-kinks \cite{Olsen,Dash}\ (localized solutions with identical boundary
conditions $\phi =0$ and $\phi =2\pi $). However, easier to follow
experimentally are non-localized $\pi $-kinks \cite{Kivshar92} (separating
regions with different values of the field $\phi ),$ see also \cite%
{ZharnitskyPRB} and references therein.

On the other hand, a \emph{breather} is spatially localized, time periodic,
oscillatory SGE--solution (see, e.g. \cite{Haskins}). It represents a field
which is periodically oscillating in time and decays exponentially in space
as the distance from the center $x=0$ is increased. This oscillatory
solution of (\ref{SGE1}) is characterized by some phase that depends on the
breather's evolution history. This could be, in particular, a bound state of
vortex with an antivortex in a Josephson junction. In this case, breather
may appear as a result of collision of a \emph{fluxon} (a propagating
magnetic flux-quantum) with an \emph{antifluxon}, or even in the process of
measurements of switching current characteristics. stationary breather
solutions form one-parameter families of solutions. An example of a
breather--solution of (\ref{SGE1}) is given by \cite{Gulevich06}:%
\begin{equation*}
\phi =4\arctan \left( \frac{\sin T}{u\cosh \left( g(u)x\right) }\right)
,\qquad
\end{equation*}%
with parameters $u=u(t)\ $and$\ T=T(t),$ such that
\begin{equation*}
g(u)=1/\sqrt{1+u^{2}}\mathrm{\quad and\quad }T(t)=\int_{0}^{t}\,g(u(t^{%
\prime }))\,u(t^{\prime })\,dt^{\prime }.
\end{equation*}%
\begin{figure}[h]
\centerline{\includegraphics[width=11.5cm]{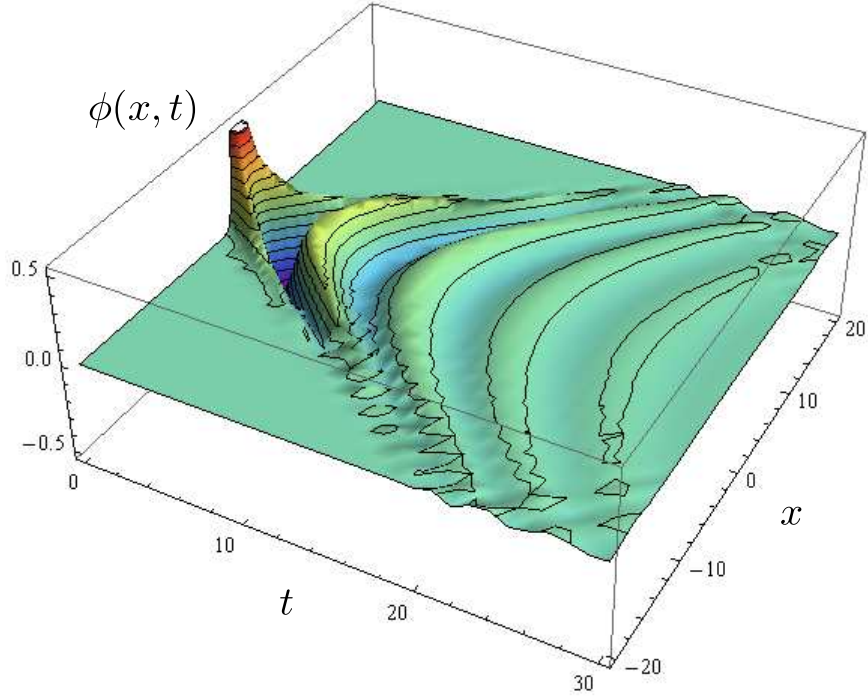}}
\caption{Numerical solution of the damped, unforced SGE (\protect\ref{sg2})
in $Mathematica$, with the following data (including the Gaussian initial
state, zero initial velocity and symmetric boundary condition): $x\in
\lbrack -20,20],~~ t\in [0,30],~~ \protect\phi (x,0)=\exp({-x^2}),~~ \protect%
\phi_{t}(x,0)=0,~~ \protect\phi (-20,t)=\protect\phi (20,t),~~\protect\gamma%
=0.2,~~F(x,t)=0$. Damping of the waves is apparent.}
\label{sgMma3}
\end{figure}

\subsubsection{Lax--pair and general SGE integrability}

In both cases (\ref{SGE1}) and (\ref{SGE2}), the SGE admits a \emph{Lax--pair%
} formulation:\footnote{%
The first Lax-pair for a nonlinear PDE was found by P. Lax in 1968
consisting of the following two operators \cite{Lax}:
\begin{equation*}
L=\frac{d^{2}}{dx^{2}}-u,\qquad M=4\frac{d^{3}}{dx^{3}}-6u\frac{d}{dx}%
-3u_{x},
\end{equation*}%
such that their Lax formulation (\ref{Lax}) gives the KdV equation:
\begin{equation*}
u_{t}-6uu_{x}+u_{xxx}=0,\qquad \text{by}
\end{equation*}%
\begin{equation*}
L_{t}=-u_{t},\quad LM-ML=u_{xxx}-6uu_{x}.
\end{equation*}%
\par
The Lax-pair form of the KdV--PDE immediately shows that the eigenvalues of $%
L$ are independent of $t$. The key importance of Lax's observation is that
any PDE that can be cast into such a framework for other operators $L$ and $%
M $, automatically obtains many of the features of the KdV--PDE, including
an infinite number of local conservation laws.}%
\begin{equation}
\dot{L}=[L,M],  \label{Lax}
\end{equation}%
where overdot means time derivative, $L$ and $M$ are linear differential
operators and $[L,M]\equiv LM-ML$ is their commutator (or, Lie bracket).
\begin{figure}[h]
\centerline{\includegraphics[width=11.5cm]{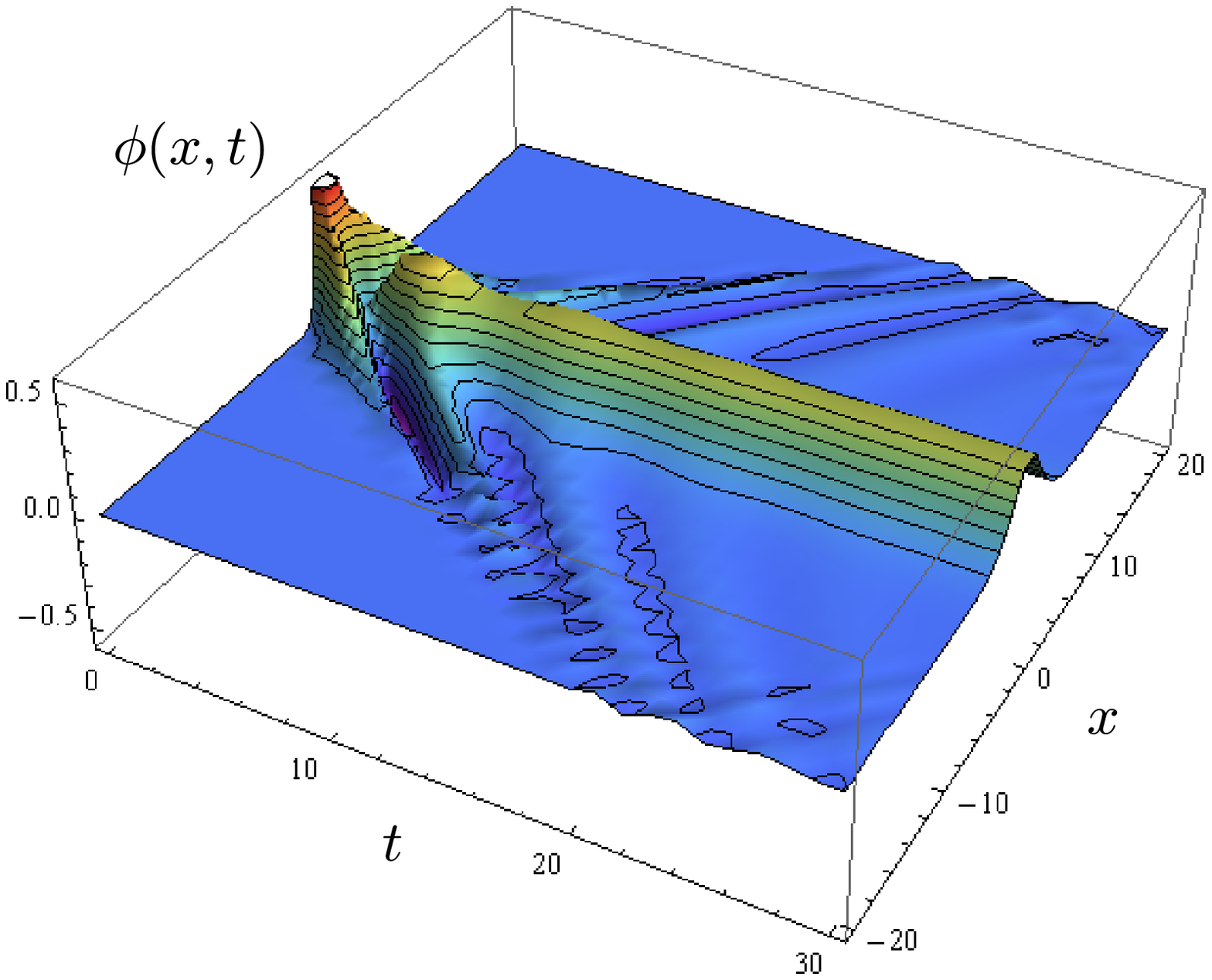}}
\caption{Numerical solution of the damped and spatially-forced SGE (\protect
\ref{sg2}) in $Mathematica$, with the following data (including the Gaussian
initial state, zero initial velocity and symmetric boundary condition): $%
x\in \lbrack -20,20],~~ t\in [0,30],~~ \protect\phi (x,0)=\exp({-x^2}),~~
\protect\phi_{t}(x,0)=0,~~ \protect\phi (-20,t)=\protect\phi (20,t),~~%
\protect\gamma=0.2,~~F(x)=0.5\func{sech}(x)$. We can see the central
sech-forcing along all time axis. Damping of the SG-waves is also apparent.}
\label{sgMma4}
\end{figure}

For example, it was shown in \cite{Li04} that the SGE (\ref{SGE1}) is
integrable through the following Lax pair:
\begin{equation}
\phi _{t}=L\phi ,\qquad \phi _{x}=M\phi ,\qquad \mathrm{where}  \label{Lax1}
\end{equation}
\begin{eqnarray*}
L &=&\left(
\begin{array}{cc}
\frac{\mathrm{i}}{4}(\phi _{x}+\phi _{t}) & -\frac{1}{16\lambda }\mathrm{e}^{%
\mathrm{i}\phi }+\lambda \\
\frac{1}{16\lambda }\mathrm{e}^{-\mathrm{i}\phi }-\lambda & -\frac{\mathrm{i}%
}{4}(\phi _{x}+\phi _{t})%
\end{array}%
\right) ,\qquad \left( \mathrm{i}=\sqrt{-1}\right) \\
M &=&\left(
\begin{array}{cc}
\frac{\mathrm{i}}{4}(\phi _{x}+\phi _{t}) & \frac{1}{16\lambda }\mathrm{e}^{%
\mathrm{i}\phi }+\lambda \\
-\frac{1}{16\lambda }\mathrm{e}^{-\mathrm{i}\phi }-\lambda & -\frac{\mathrm{i%
}}{4}(\phi _{x}+\phi _{t})%
\end{array}%
\right) ,\qquad \left( \lambda \in \mathbb{R}\right) .
\end{eqnarray*}%
The Lax pair (\ref{Lax1}) possesses the following complex-conjugate
symmetry: if $\phi =\left(
\begin{array}{c}
\phi _{1} \\
\phi _{2}%
\end{array}%
\right) $ solves the Lax pair (\ref{Lax1}) at $(\lambda ,\phi )$, then $%
\left(
\begin{array}{c}
\overline{\phi _{2}} \\
\overline{\phi _{1}}%
\end{array}%
\right) $ solves the Lax pair (\ref{Lax1}) at $(-\bar{\lambda},\phi )$. In
addition, there is a Darboux transformation for the Lax pair (\ref{Lax1}) as
follows: let
\begin{equation*}
u=\phi +2\mathrm{i}\ln \left[ \frac{\mathrm{i}\phi _{2}}{\phi _{1}}\right]
,\qquad \left( u\in \mathbb{R}\right) .\qquad
\end{equation*}%
If $\phi =\phi |_{\lambda =\nu }$ for some $\nu \in \mathbb{R},$\ then
\begin{equation*}
\psi =\left(
\begin{array}{cc}
-\nu \phi _{2}/\phi _{1} & \lambda \\
-\lambda & \nu \phi _{1}/\phi _{2}%
\end{array}%
\right) \phi .
\end{equation*}%
solves the Lax pair (\ref{Lax1}) at $(\lambda ,u)$. Also, from its spatial
part: $\phi _{x}=M\phi $, a complete Floquet theory can be developed. See
\cite{Li04} for the proofs and more technical details.
\begin{figure}[h]
\centerline{\includegraphics[width=11.5cm]{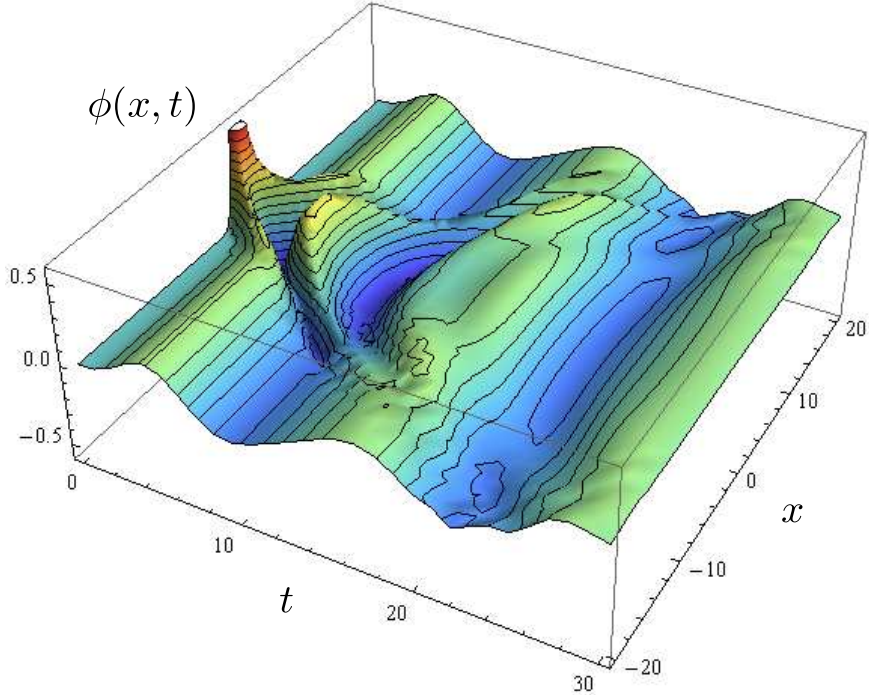}}
\caption{Numerical solution of the damped and temporally-forced SGE (\protect
\ref{sg2}) in $Mathematica$, with the following data (including the Gaussian
initial state, zero initial velocity and symmetric boundary condition): $%
x\in \lbrack -20,20],~~ t\in [0,30],~~ \protect\phi (x,0)=\exp({-x^2}),~~
\protect\phi_{t}(x,0)=0,~~ \protect\phi (-20,t)=\protect\phi (20,t),~~%
\protect\gamma=0.2,~~F(x)=0.1\func{sin}(t/2)$. We can see the sine-forcing
along all time axis. Damping of the SG-waves is also apparent.}
\label{sgMma5}
\end{figure}

\subsection{SGE modifications}

\subsubsection{SGE with the positive sine term}

The simplest SGE modification is to replace the minus sine term with the
plus sine:
\begin{equation}
\phi _{tt}=\phi _{xx}+\sin \phi ,\qquad \text{\textrm{or}}\qquad \phi
_{tt}(x,t)=\phi _{xx}(x,t)+\sin \phi (x,t).  \label{SGE+}
\end{equation}%
Again, a typical, spatially-symmetric, boundary-value problem for (\ref{SGE+}%
) is defined by:%
\begin{eqnarray*}
x &\in &[-L,L]\subset \mathbb{R},\qquad \left( t\in \mathbb{R}^{+}\right) ,%
\hspace{1.7cm} \\
\phi (x,0) &=&f(x),\quad \phi _{t}(x,0)=0,\quad \phi (-L,t)=\phi (L,t),
\end{eqnarray*}%
where $f(x)\in \mathbb{R}$ is an axially-symmetric function (see Figure \ref%
{sgMma2}).

\subsubsection{Perturbed SGE and $\protect\pi$--kinks}
\label{pertSG}

As we have seen above (and it was proved by \cite{Ablowitz81,Rajaraman}),
the (1+1) SGE is integrable.~In general though, the perturbations to this
equation associated with the external forces and inhomogeneities spoil its
integrability and the equation can not be solved exactly. Nevertheless, if
the influence of these perturbations is small, the solution can be found
perturbatively \cite{Gulevich06}. The perturbation theory for solitons was
developed by \cite{McLaughlin77} and subsequently applied by \cite%
{McLaughlin78} to dynamics of vortices in Josephson contacts. Perturbed SGEs
come in a variety of forms (see, e.g. \cite%
{McLaughlin77,McLaughlin78,Kaup78,Kaup90}).
\begin{figure}[h]
\centerline{\includegraphics[width=11.5cm]{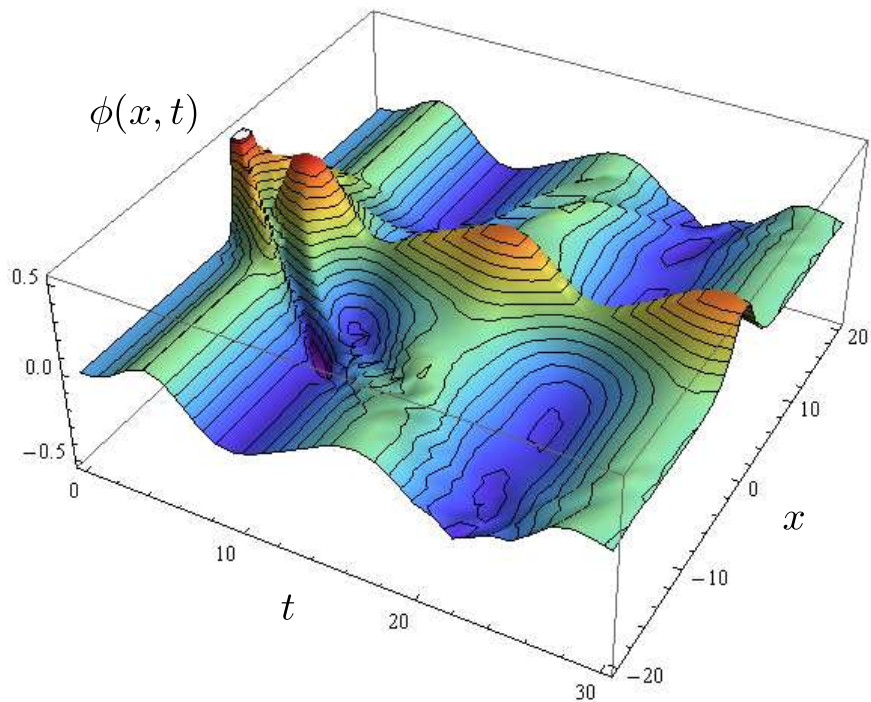}}
\caption{Numerical solution of the damped and both spatially and temporally
forced SGE (\protect\ref{sg2}) in $Mathematica$, with the following data
(including the Gaussian initial state, zero initial velocity and symmetric
boundary condition): $x\in \lbrack -20,20],~~ t\in [0,30],~~ \protect\phi %
(x,0)=\exp({-x^2}),~~ \protect\phi_{t}(x,0)=0,~~ \protect\phi (-20,t)=%
\protect\phi (20,t),~~\protect\gamma=0.2,~~F(x,t)=0.1\func{sin}(t/2)+0.5%
\func{sech}(x)$. We can see both temporal sine-forcing and spatial
sech-forcing along all time axis. Damping of the SG-waves is still visible.}
\label{sgMma6}
\end{figure}

One common form is a damped and driven SGE:%
\begin{equation}
\phi _{tt}+\gamma \phi _{t}-\phi _{xx}+\sin \phi =F,  \label{sg2}
\end{equation}%
where $\gamma \phi _{t}$ is the damping term and $F(x,t)$ is the
spatiotemporal forcing. Special cases of the forcing term $F=F(x,t)$ in (\ref%
{sg2})\ are: (i) purely temporal $F=F(t)$ (e.g., periodic, see Figure \ref%
{sgMma4}); (ii) purely spatial $F=F(x)$ (e.g., central-symmetric, see Figure %
\ref{sgMma5}); and (iii) spatiotemporal $F=F(x,t)$ (e.g.,
temporally-periodic and spatially central-symmetric, see Figure \ref{sgMma6}%
).

Considering (for simplicity) purely spatial forcing: $F(x,t)=F(x)$, it has
been shown in \cite{GonzalezPre,GonzalezCha} that if $F(x_{0})=0$ for some
point $x_{0}\in \mathbb{R}$, this can be an equilibrium position for the
soliton. If there is only one zero, in case of a soliton this is a stable
equilibrium position if $\left( \frac{\partial F(x)}{\partial x}\right)
_{x_{0}}>0$; in case of an antisoliton, this is a stable equilibrium
position if $\left( \frac{\partial F(x)}{\partial x}\right) _{x_{0}}<0$ (see
and references therein).

In particular if%
\begin{equation*}
F(x)=2(\beta ^{2}-1)\sinh (\beta x)/\cosh ^{2}(\beta x),\qquad (\beta \in
\mathbb{R}),
\end{equation*}%
the exact stationary kink--solution of (\ref{sg2}) is:
\begin{equation*}
\phi _{k}=4\arctan \left[ \exp \left( \beta x\right) \right] .
\end{equation*}%
The stability analysis, which considers small amplitude oscillations around%
\newline
\ $\phi _{k}\left[ \phi (k,x)=\phi _{k}(x)+f(x)\mathrm{e}^{\lambda t}\right]
,$ leads to the following eigenvalue problem:
\begin{equation*}
\widehat{L}f=\Gamma f,\quad \mathrm{where\quad }\ \widehat{L}=-\partial
_{x}^{2}+\left[ 1-2\cosh ^{-2}(\beta x)\right] \quad \mathrm{and\quad }%
\Gamma =-\lambda ^{2}-\gamma \lambda \;.
\end{equation*}%
The eigenvalues of the discrete spectrum are given by the formula
\begin{equation*}
\Gamma _{n}=\beta ^{2}(\Lambda +2\Lambda n-n^{2})-1,
\end{equation*}%
where $\Lambda (\Lambda +1)=2/\beta ^{2}$. The integer part of $\Lambda $,
yields the number of eigenvalues in the discrete spectrum, which correspond
to the soliton modes (this includes the translational mode $\Gamma _{0}$,
and the internal or shape modes $\Gamma _{n}$ with $n>0$ (see \cite%
{GonzalezPre,GonzalezCha}).

In case of a function $F$ defined in such a way that it possesses many
zeroes, maxima and minima, perturbed SGE (\ref{sg2}) describes an array of
inhomogeneities. For example,%
\begin{equation*}
F(x)={\sum_{n=-q}^{q}4\left( 1-\beta ^{2}\right) \frac{\mathrm{e}^{\beta
\left( x+x_{n}\right) }-\mathrm{e}^{3\beta \left( x+x_{n}\right) }}{\left(
\mathrm{e}^{2\beta \left( x+x_{n}\right) }+1\right) ^{2}},}
\end{equation*}%
where $x_{n}=(n+2)\log \left( \sqrt{2}+1\right) /\beta $ ($n=-q,-q+1\cdots
,q-1,q$), and $q+2$ is the number of extrema points of $F(x)$. When the
soliton is moving over intervals where $\frac{dF(x)}{dx}<0$, its internal
mode can be excited. The points $x_{i}$ where $F(x_{i})=0$ and $\frac{%
dF(x_{i})}{dx}<0$, are `barriers'\ which the soliton can overcome due to its
kinetic energy (for more details, see \cite{GonzalezPre,GonzalezCha}).

Study of non-localized $\pi $-kinks in parametrically forced SGE (PSGE):%
\begin{equation}
\phi _{tt}\;=\;\phi _{xx}\;-\;a(t/\epsilon )\,\sin \phi ,  \label{igor9}
\end{equation}%
(over the fast time scale $\epsilon ,$ where $a$ is a mean-zero periodic
function with a unit amplitude), has been performed by \cite%
{ZharnitskyPRB,ZharnitskyPRE}, via $2\pi $-kinks as approximate solutions.
In particular, a finite-dimensional counterpart of the phenomenon of $\pi $%
-kinks in PSGE is the stabilization of the inverted \emph{Kapitza pendulum}
by periodic vibration of its suspension point. Geometrical averaging
technique\footnote{%
The averaged forces in a rapidly forced system (e.g. inverted \emph{Kapitza
pendulum}) are the constraint forces of an associated auxiliary
non-holonomic system; the curvature of these constraints enters the
expression for the averaged system \cite{Levi}.} of \cite{Levi} was applied
as a series of canonical near-identical transformations via Arnold's normal
form technique \cite{Arnold}, as follows.

Starting with the Hamiltonian of PSGE (\ref{igor9}), given by:
\begin{equation*}
H(\phi )=\int_{-\infty }^{+\infty }\left( \frac{p^{2}}{2}+\frac{\phi _{x}^{2}%
}{2}-a\cos {\phi }\right) dx,\qquad \mathrm{where}\text{ \ }\left( p\,\equiv
\,\phi _{t}\equiv \dot{\phi}\right) ,
\end{equation*}%
a series of canonical transformations was performed in \cite{ZharnitskyPRB}
with the aim to kill all rapidly-oscillating terms, the following
slightly-perturbed Hamiltonian was obtained:
\begin{equation*}
H_{\mathrm{per}}=\int_{-\infty }^{+\infty }\left( \frac{p_{3}^{2}}{2}+\frac{%
\phi _{3x}^{2}}{2}+\frac{1}{2}\epsilon ^{2}\langle a_{-1}^{2}\rangle \sin
^{2}{\phi _{3}}\right) dx+O(\epsilon ^{3}),
\end{equation*}%
which, after rescaling: $X=\epsilon x,\;T=\epsilon t,\;P=2\epsilon
^{-1}p_{3},\;\Phi =2\phi _{3},$ gave the following system of a slightly
perturbed SGE with $2\pi $-kinks as approximate solutions:%
\begin{equation*}
\Phi _{T}=P+O(\epsilon ^{2}),\qquad P_{T}=\Phi _{XX}-\langle
a_{-1}^{2}\rangle \sin {\Phi }+O(\epsilon ),
\end{equation*}%
where $a_{-1}$ is an anti-derivative with zero average. Finally, after
rescaling back to variables $(\phi _{3},p_{3})$, approximate solutions $\phi
_{3}\approx \psi (x,t)$ in the form of $\pi $-kinks were obtained, with
\begin{equation*}
\psi (x,t)=2\arctan {\left[ \exp {\left( \epsilon \sqrt{\langle
a_{-1}^{2}\rangle }\frac{x-ct}{\sqrt{1-c^{2}}}\right) }\right] },
\end{equation*}%
where $c$ is the wave-propagation velocity. For more technical details, see
\cite{ZharnitskyPRB}.

In addition, he following two versions of the perturbed SGE have been
studied in \cite{ZharnitskyPRE}:
\begin{enumerate}
\item Directly forced SGE:
\begin{equation*}
\phi _{tt}-\phi _{xx}+\sin {\phi }=Mf(\omega t).
\end{equation*}%
After shifting to the oscillating reference frame by the transformation:%
\begin{equation}
\phi =\theta +M\omega ^{-2}F(\omega t),  \label{vadim1}
\end{equation}%
where $F$ has zero mean and $F^{\prime \prime }(\tau )=f(\tau ),$ the
parametrically forced ODE is obtained:
\begin{eqnarray}
\ddot{\theta} &=&-\sin {(\theta +M\omega ^{-2}F(\omega t))}\,,\qquad \mathrm{%
with}  \label{odeth} \\
H &=&\frac{p^{2}}{2}-A(\omega t)\cos (\theta )+B(\omega t)\sin (\theta ),
\notag
\end{eqnarray}
where $p$ is the momentum canonically conjugate to $\theta $, and
\begin{equation*}
A(\omega t)=\cos (M\omega ^{-2}F(\omega t)),\qquad B(\omega t)=\sin (M\omega
^{-2}F(\omega t)).
\end{equation*}%
From (\ref{odeth}), the corresponding evolution PDE (in canonical form) is
obtained for a new phase $\theta $ on top of a rapidly oscillating
background field:%
\begin{equation*}
\theta _{t}=p,\qquad p_{t}=\theta _{xx}-\sin {(\theta +M\omega ^{-2}F(\omega
t))}.
\end{equation*}%
After retracing the identical transformation (\ref{vadim1}), the so-obtained
(approximate) solutions become $\pi $-kinks (see \cite{ZharnitskyPRE} for
technical details).

\item Damped and driven SGE
\begin{equation}
\phi _{tt}-\phi _{xx}+\sin {\phi }=Mf(\omega t)-\alpha \phi _{t}+\eta ,
\label{dfdsge}
\end{equation}%
which is frequently used to describe long Josephson junctions \cite%
{McLaughlin78}.\footnote{%
In (\ref{dfdsge}), $\phi $ represents the phase-difference between the
quantum-mechanical wave functions of the two superconductors defining the
Josephson junction, $t$ is the normalized time measured relative to the
inverse plasma frequency, $x$ is space normalized to the Josephson
penetration depth, while $Mf(\omega t)$ represents tunneling of
superconducting Cooper pairs (normalized to the critical current density).}
Starting with a homogeneous transformation to the oscillating reference
frame, analogous to~(\ref{vadim1}) and designed to remove the free
oscillatory term: $\phi =\theta +G(t),$ and substituting this transformation
to (\ref{dfdsge}), while choosing the function $G$ so that it solves the ODE
\begin{equation*}
\ddot{G}+\alpha \dot{G}=Mf(\omega t),
\end{equation*}
the following evolution PDE is obtined (in canonical form) \cite%
{ZharnitskyPRE}:
\begin{equation}
\theta _{t}=p,\qquad p_{t}=\theta _{xx}-\alpha p+\eta -\sin (\theta
+G(\omega t)).  \label{efeq}
\end{equation}%
For the particular case of $f(\tau )=\sin {\tau }$, the the function $G$ is
found to be:
\begin{equation*}
G(\tau )=-\frac{\alpha }{\omega }\frac{M}{\alpha ^{2}+\omega ^{2}}\cos {\tau
}-\frac{M}{\alpha ^{2}+\omega ^{2}}\sin {\tau }.
\end{equation*}%
After a series of transformations (related to a directly-forced pendulum),
in zeroth order in $\alpha ,\eta $, the evolution PDE (\ref{efeq}) reduces
(after neglecting terms $\sim \omega ^{-3}$) to SGE, which has $\pi $-kink
solutions. Therefore, slightly perturbed $\pi $-kinks are approximate
solutions of the original equation (\ref{dfdsge}) (on top of the rapidly
oscillating background field; see \cite{ZharnitskyPRE} for technical
details).
\end{enumerate}

\subsubsection{SGE in (2+1) dimensions}

The (2+1)D SGE with additional spatial coordinate ($y$) is defined on $%
\mathbb{R}^{2,1}$as:\footnote{%
In the case of a long Josephson junction, the soliton solutions of (\ref%
{2DSG}) describe Josephson vortices or \emph{fluxons}. These excitations are
associated with the distortion of a Josephson vortex line and their shapes
can have an arbitrary profile, which is retained when propagating. In (\ref%
{2DSG}), $\varphi $ denotes the superconducting phase difference across the
Josephson junction; the coordinates $x$ and $y$ are normalized by the
Josephson penetration length $\lambda _{J}$, and the time $t$ is normalized
by the inverse Josephson plasma frequency $\omega _{p}^{-1}$~(see \cite%
{Gulevich08} and references therein).}%
\begin{equation}
\varphi _{tt}=\Delta \varphi -\sin \varphi =\varphi _{xx}+\varphi _{yy}-\sin
\varphi .  \label{2DSG}
\end{equation}

A special class of solutions of (\ref{2DSG}) can be constructed by
generalization of the solution of (\ref{SGE1}) which does not depend on one
of the coordinates, or, obtained by Lorentz transforming the solutions of a
stationary 2D SGE. However, there are numerical solutions of (\ref{2DSG})
which cannot be derived from the (\ref{SGE1}) or (\ref{SGE2}), e.g., radial
breathers (or, pulsons) \cite{Gulevich08}.

A more general class of solutions of the (2+1)D SG equation has the
following form,\footnote{%
Because of the arbitrariness of $f$, solution~(\ref{phi}) describes a
variety of excitations of various shapes. Choosing~$f$ localized in a finite
area, e.g., $f=A/\cosh (x-t)$, solution~(\ref{phi}) describes an excitation,
localized along $x$ that keeps its shape when propagating, i.e., a solitary
wave (in the sense of~\cite{Rajaraman}). For each solitary wave of this
type, there exists an anti-partner with an $f$ of opposite sign in (\ref{phi}%
). For solitary waves to be solitons, there is an additional important
criterion: restoring their shapes after they collide.
\par
Consider a trial function
\begin{equation*}
\varphi (x,y,t)=4\,\arctan \exp \left[ y-f(x+t)\pm f(x-t)\right] ,
\end{equation*}%
that, when $t\rightarrow -\infty $, describes the propagation of two
solitary shape waves toward each other (minus sign) or a solitary wave and
its anti-partner (plus sign). One can see that~(\ref{phi}) can only \textit{%
approximately} satisfy (\ref{2DSG}) when $|f^{\prime }(x+t)f^{\prime
}(x-t)|\ll 1$ for all values of $x$ and $t$. This suggests that, in general,
the condition for restoring the shapes may not be satisfied. In general
case, (\ref{2DSG}) can not be satisfied, that prompts that the collision of
two solitary waves leads to distortion of the original excitations \cite%
{Gulevich08}.}
\begin{equation}
\varphi (x,y,t)=4\,\arctan \exp \left[ y-f(x\pm t)\right] ,  \label{phi}
\end{equation}%
which exactly satisfies (\ref{2DSG}) with an arbitrary real-valued
twice-differentiable function $f=f(x\pm t)$. The excitations, described by $%
f $ are similar to elastic shear waves in solid mechanics~\cite{shear-waves}.

Since the equation~(\ref{2DSG}) is Lorentz-covariant, we can obtain other
solutions performing Lorentz transformations on (\ref{phi}), which leads to
a class of solutions of the form \cite{Gulevich08}:
\begin{equation*}
\varphi (x,y,t)=4\,\arctan \exp \left[ \frac{y-v\,t}{\sqrt{1-v^{2}}}-f\left(
x\pm \frac{t-v\,y}{\sqrt{1-v^{2}}}\right) \right] .
\end{equation*}

\subsubsection{Two coupled SGEs}

\begin{figure}[h]
\centerline{\includegraphics[width=12cm]{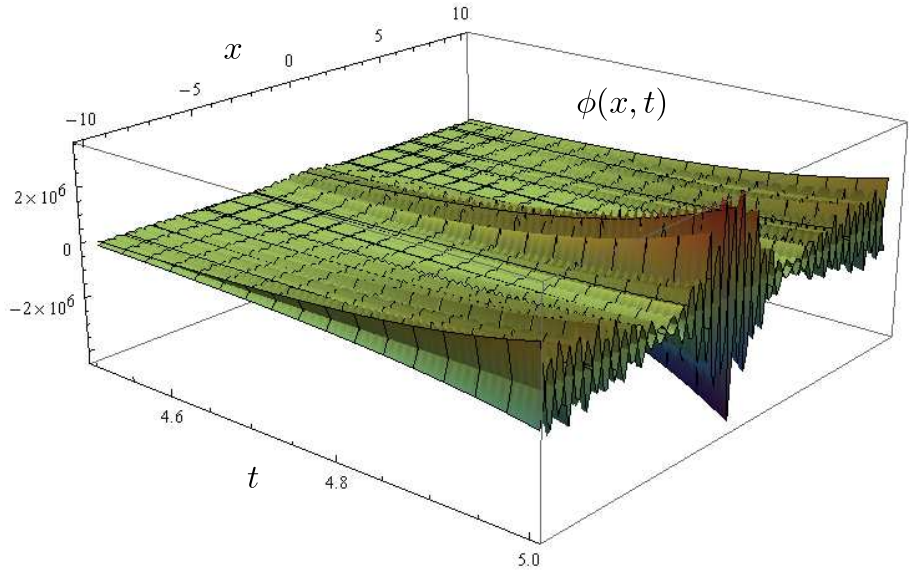}}
\caption{Numerical solution of the SGE-system (\protect\ref{2SGE}) in $%
Mathematica$, with the following data: ~$x\in \lbrack -10,10],~~ t\in
[0,5],~~\protect\alpha =0.5,~~\protect\beta=0.3,~~ \protect\phi(x,0)=0.3\exp(%
{-x^2}),~~\protect\psi(x,0)=0.7\exp({-x^2}),~~ \protect\phi_{t}(x,0)=0,~~%
\protect\phi(-10,t)=\protect\phi(10,t), ~~\protect\psi_{t}(x,0)=0,~~\protect%
\psi(-10,t)=\protect\psi(10,t)$.}
\label{2SG}
\end{figure}

The following two-parameter system of two coupled SGEs was introduced by
\cite{Khusnutdinova}:
\begin{eqnarray}
\phi _{tt}-\phi _{xx} &=&-\beta ^{2}\sin (\phi -\psi ),  \label{2SGE} \\
\psi _{tt}-\alpha ^{2}\psi _{xx} &=&\sin (\phi -\psi ),\qquad \text{\textrm{%
with constants} }(\alpha ,\beta >0).  \notag
\end{eqnarray}%
For numerical solution, see Figure \ref{2SG}.

The SGE-system (\ref{2SGE}) has been exactly solved by \cite{Salas10}, where
(using a series of substitutions) it was first reduced to the nonlinear
second-order ODE:
\begin{equation}
\varphi ^{\prime \prime }(\xi )=\left[ \frac{1+\alpha ^{2}\beta
^{2}-c^{2}\left( 1+\beta ^{2}\right) }{\left( c^{2}-1\right) \left(
c^{2}-\alpha ^{2}\right) \mu ^{2}}+\frac{2\varphi ^{\prime 2}}{\varphi (\xi
)^{2}+1}\right] \varphi (\xi ),  \label{vx}
\end{equation}%
equivalent to the following autonomous system in the $\left( X,Y\right)-$%
plane:%
\begin{equation}
\frac{dX}{d\xi }=Y,\qquad \frac{dY}{d\xi }=\left[ \frac{1+\alpha ^{2}\beta
^{2}-c^{2}\left( 1+\beta ^{2}\right) }{\left( c^{2}-1\right) \left(
c^{2}-\alpha ^{2}\right) \mu ^{2}}+\frac{2Y^{2}}{X^{2}+1}\right] X.
\label{X-Y}
\end{equation}%
System (\ref{X-Y}) has an equilibrium point at the origin: $\left(
X,Y\right) =(0,0),$ in which the Jacobian matrix is:%
\begin{equation*}
J_{(0,0)}=\left(
\begin{array}{cc}
0 & 1 \\
\lambda & 0%
\end{array}%
\right) ,\qquad \text{with}\qquad \lambda =\frac{\alpha ^{2}\beta
^{2}-c^{2}\left( 1+\beta ^{2}\right) +1}{\left( c^{2}-1\right) \left(
c^{2}-\alpha ^{2}\right) \mu ^{2}}.
\end{equation*}%
The phase portraits from this system show that there exist periodic
solutions of the coupled SGEs (\ref{2SGE}).

Using the exponential ansatz:
\begin{equation*}
\varphi (\xi )=\frac{p\exp (\xi )+q\exp (-\xi )}{r\exp (\xi )+s\exp (-\xi )}%
,\qquad \text{\textrm{with constants} }(p,q,r,s\in \mathbb{R}),
\end{equation*}%
four pairs of analytic solutions of the system (\ref{vx}), and therefore of
the system (\ref{2SGE}), were found in \cite{Salas10}. We present here only
the first two (simpler) solution pairs:
\begin{eqnarray*}
\phi_1(x,t) &=&\frac{\beta ^{2}}{4\left( c^{2}-1\right) \mu ^{2}}\sin (2\xi
)+c_{1}\xi +c2,\quad \\
\psi_1(x,t) &=&\frac{\beta ^{2}}{4\left( c^{2}-1\right) \mu ^{2}}\sin (2\xi
)+c_{1}\xi +c2-2\arctan (\tan (\xi )), \\
\xi &=&\mu (x-ct),\,\qquad c=\sqrt{\frac{1-\alpha ^{2}\beta ^{2}}{1-\beta
^{2}}}.
\end{eqnarray*}
\begin{eqnarray*}
\phi_2(x,t) &=&\frac{\beta ^{2}}{4\left( c^{2}-1\right) \mu ^{2}}\sin (2\xi
)+c_{1}\xi +c2,\quad \\
\psi_2(x,t) &=&\frac{\beta ^{2}}{4\left( c^{2}-1\right) \mu ^{2}}\sin (2\xi
)+c_{1}\xi +c2-2\arctan (\cot (\xi )), \\
\xi &=&\mu (x-ct),\,\qquad c=\sqrt{\frac{1-\alpha ^{2}\beta ^{2}}{1-\beta
^{2}}}.
\end{eqnarray*}

For more technical details, see \cite{Salas10}.

\subsection{Sine--Gordon chain and discrete breathers}

\subsubsection{Frenkel--Kontorova model}

The original Frenkel--Kontorova model \cite{FrenKont,Braun98,Braun04} of
stationary and moving crystal dislocations, was formulated historically
decades before the continuous SGE. It consists of a chain of harmonically
coupled atoms in a spatially periodic potential, governed by the set of
differential-difference equations:
\begin{equation}
\ddot{\phi}_{n}+\frac{1}{\Delta x^{2}}\left[ \phi _{n+1}-2\phi _{n}+\phi
_{n-1}\right] +\sin \phi _{n}=0,\,  \label{FK}
\end{equation}%
where $\phi _{n}$ denotes the position of the $n$th atom in the chain.
Alternatively, system (\ref{FK}) represents a chain of torsionally-coupled
pendula (see Figure \ref{solitPendula}), where $\phi _{n}$ is the angle
which the $n$th pendulum makes with the vertical.

\subsubsection{Sine--Gordon chain}

To derive dynamical equations of the sine--Gordon chain (SGC), consisting of
anharmonic oscillators with the coupling constant $\mu $, we start with the
three-point, central, finite-difference approximation of the spatial
derivative term $\phi _{xx}$\ in the SGE:
\begin{eqnarray*}
\phi _{xx} &\approx &\frac{1}{\Delta x^{2}}\left[ \phi _{n+1}-2\phi
_{n}+\phi _{n-1}\right] +O(x^{2}) \\
&=&-\frac{1}{\Delta x^{2}}\left[ (\phi _{n}-\phi _{n-1})-(\phi _{n+1}-\phi
_{n})\right] +O(x^{2}).
\end{eqnarray*}%
Applying this finite-difference approximation to the SGE (\ref{SGE1}), and
also performing the corresponding replacements: $\phi \rightarrow \phi
_{n},~\phi _{tt}\rightarrow \ddot{\phi}_{n}$ and $\mu =1/\Delta x^{2},$ we
obtain the set of difference ODEs defining the SGC:%
\begin{equation}
\ddot{\phi}_{n}+\mu \left[ (\phi _{n}-\phi _{n-1})-(\phi _{n+1}-\phi _{n})%
\right] +\sin \phi _{n}=0\,.  \label{SGC}
\end{equation}%
The system (\ref{SGC}) describes a chain of interacting particles subjected
to a periodic on-site potential $V(x)=\sin (x)$. In the continuum limit, (%
\ref{SGC}) becomes the standard SGE (\ref{SGE1}) and supports stable
propagation of a kink-soliton of the form (\ref{kink1}).
\begin{figure}[h]
\centerline{\includegraphics[width=9cm]{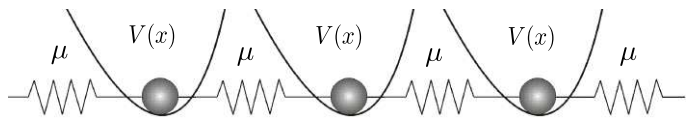}}
\caption{Simple sine--Gordon chain (SGC) with the coupling constant $\protect%
\mu $ and the periodic on-site potential $V(x)=\sin (x)$.}
\label{chain}
\end{figure}

The linear-wave spectrum of (\ref{SGC}) around a kink has either one or two
localized modes (which depends on the value of $\mu )$ \cite{Prilepsky04}.
The frequencies of these modes lie inside the spectrum gap. The linear
spectrum, with the linear frequency $\omega $ and the wave number $k,$ is
given by:
\begin{equation}
\omega ^{2}=1+4\mu \sin ^{2}\frac{k}{2},  \label{spec}
\end{equation}
while the gap edge frequency is $\omega =1.$

The simplest example of (\ref{SGC}), containing only two oscillators, is
defined by \cite{Prilepsky04}:
\begin{equation}
\ddot{\phi}_{1,\,2}+\mu (\phi _{1,\,2}-\phi _{2,\,1})+\sin \phi _{1,\,2}=0\,.
\label{2p}
\end{equation}

It was demonstrated in \cite{Jensen92}, using the method of averaging in
fast oscillations, that a perturbed SGC, damped and driven by a
large-amplitude ac-force, might support localized kink solitons.
Specifically, they considered the perturbed SGC:
\begin{equation}
\ddot{\phi}_{n}-\mu \left[ \phi _{n+1}-2\phi _{n}+\phi _{n-1}\right] +\sin
\phi _{n}=\chi +\alpha \sin \omega t-\gamma \dot{\phi}_{n},  \label{SGC2}
\end{equation}%
where $\chi $ is a dc-force, $\alpha $ and $\omega $ are the normalized
(large) amplitude and frequency of a periodic force, respectively, while $%
\gamma $ is the normalized dissipative coefficient. Without the forcing on
the right-hand side, (\ref{SGC2}) reduces to (\ref{SGC}).

\subsubsection{Continuum limits}

Perturbed SGEs have their corresponding perturbed SGCs. The following $0$-$%
\pi ~$SGC was proposed in \cite{Derks}:
\begin{equation}
\ddot{\phi _{n}}=\frac{\phi _{n-1}-2\phi _{n}+\phi _{n+1}}{a^{2}}-\sin (\phi
_{n}+\theta _{n})+\gamma ,  \label{dsg}
\end{equation}%
as an equation of a phase $\phi _{n}$-motion (of a $0$-$\pi $ array of
Josephson junctions). Here, $a$ is the lattice spacing parameter, $\gamma >0$
is the applied bias current density, and $\theta _{n}=(0$ if $n\leq 0$ and $%
-\pi $ if $n>0)$ is the phase jump of $\pi $ in $\phi _{n}$. The SGC
equation (\ref{dsg}) is derived from the following discrete Lagrangian:%
\begin{equation}
L_{\mathrm{D}}=\int {\sum_{n\in \mathbb{Z}}{\left[ \frac{1}{2}\left( \frac{%
d\phi _{n}}{dt}\right) ^{2}-\frac{1}{2}\left( \frac{\phi _{n+1}-\phi _{n}}{a}%
\right) ^{2}-1+\cos (\phi _{n}+\theta _{n})+\gamma \phi _{n}\right] }\,dt}.
\label{lag1}
\end{equation}%
In the continuum limit $a\ll 1$ Lagrangian (\ref{lag1}) becomes
\begin{equation*}
L_{\mathrm{C}}=\iint_{-\infty }^{\infty }\left[ \,\frac{1}{2}\left( \phi
_{t}\right) ^{2}-\frac{1}{2}\left( \widetilde{L}_{a}\phi _{x}\right)
^{2}-1+\cos (\phi +\theta )+\gamma \phi \right] \,dx\,dt\,,
\end{equation*}%
from which, the continuum limit of (\ref{dsg}) gives the following perturbed
SGE:%
\begin{equation*}
\phi _{tt}=L_{a}\phi _{xx}-\sin (\phi +\theta )+\gamma ,
\end{equation*}%
where $\theta =(0$ if $x\leq 0$ and $-\pi $ if $x>0)$, while the
differential operators $L_{a}\phi _{xx}$ and $\widetilde{L}_{a}\phi _{x}$
are given by the following Taylor expansions:
\begin{eqnarray*}
L_{a}\phi _{xx} &=&\frac{\phi _{n-1}-2\phi _{n}+\phi _{n+1}}{a^{2}}%
=2\sum_{k=0}^{\infty }\frac{a^{2k}}{(2k+2)!}\,\partial _{xx}^{k}\phi
_{xx}(na), \\
\widetilde{L}_{a}\phi _{x} &=&\frac{\phi _{n+1}-\phi _{n}}{a}%
=\sum_{k=0}^{\infty }\frac{a^{k}}{(k+1)!}\,\partial _{x}^{k}\phi (na).
\end{eqnarray*}%
For more technical details, including several other continuum limits, see
\cite{Derks}.

\subsubsection{Discrete breathers}

\begin{figure}[h]
\centerline{\includegraphics[width=9cm]{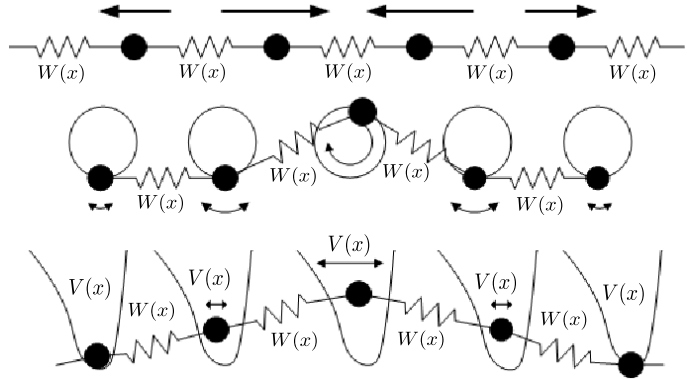}}
\caption{Different types of discrete breathers (DBs): acoustic breather
(top), rotobreather (middle), and optical breather (bottom); modified and
adapted from \protect\cite{Aubry,Flach03,FlachRep,Flach05}).}
\label{DB}
\end{figure}

Generally speaking, it is a well-known fact (see, e.g. \cite{Flach08} and
references therein) that different types of \emph{excitations}, most notably
\emph{phonons} (propagating linear waves) and \emph{discrete breathers} (DBs
for short; they are time-periodic spatially localized excitations, also
labeled intrinsic localized modes or discrete solitons) can occur as
solutions of spatially-discrete nonlinear lattices. According to S. Flach
\emph{et al.} \cite{FlachRep,Flach03,Flach08} DBs are caused by a specific
interplay between the nonlinearity and discreteness of the lattice. The
lattice nonlinearity provides with an amplitude-dependent tunability of
oscillation or rotation frequencies of DBs, while its spatial discreteness
leads to finite upper bounds of the frequency spectrum of small amplitude
waves.\footnote{%
DBs are not sensitive to specific types of nonlinearities in the lattice nor
are they confined to any lattice dimensions; they are (usually) dynamically
and structurally stable and emerge in a variety of physical systems (ranging
from lattice vibrations and magnetic excitations in crystals to light
propagation in photonic structures and cold atom dynamics in periodic
optical traps, see \cite{FlachSpr}).} Although DBs present complex dynamical
objects, experimental measurements can (in many cases) be well understood by
using their \textit{time-averaged} properties (see \cite{aemsfmvfyzjbp01}).
In addition, nonlinear discrete lattices admit different types of DBs
depending on the spectrum of linear waves propagating in the lattice \cite%
{Aubry,Flach03,FlachRep,Flach05}, including: acoustic breathers,
rotobreathers and optical breathers (see Figure \ref{DB}).

A particular system studied in \cite{Flach03,FlachSpr} has been
characterized by the lattice Hamiltonian:%
\begin{eqnarray}
H &=&\sum_{n}(\frac{1}{2}\dot{x}_{n}^{2}+W(x_{n}-x_{n-1})+V(x_{n}))
\label{hlat} \\
&=&\sum_{n}(\frac{1}{2}p_{n}^{2}+W(x_{n}-x_{n-1})+V(x_{n})),  \notag
\end{eqnarray}%
where $x_{n}=x_{n}(t)$ are time-dependent coordinates with
canonically-conjugate momenta $p_{n}=\dot{x}_{n}(t),$ $W(x_{n})=W(x)$ is the
nearest neighbor interaction, and $V(x_{n})=V(x)$ is an optional on-site
(substrate) potential. From (\ref{hlat}) the following equations of motion
are derived:%
\begin{eqnarray*}
\ddot{x}_{n} &=&-W^{\prime }(x_{n}-x_{n-1})+W^{\prime
}(x_{n+1}-x_{n})-V^{\prime }(x_{n}),\qquad \text{\textrm{or}} \\
\dot{x}_{n} &=&\dot{x}_{n},\qquad \dot{p}_{n}=-W^{\prime
}(x_{n}-x_{n-1})+W^{\prime }(x_{n+1}-x_{n})-V^{\prime }(x_{n}),
\end{eqnarray*}%
where (for simplicity) the following zero initial conditions are assumed:
\begin{equation*}
V(0)=W(0)=V^{\prime }(0)=W^{\prime }(0)=0,~~~V^{\prime \prime }(0)\geq
0,~~~W^{\prime \prime }(0)>0.
\end{equation*}%
Hamiltonian (\ref{hlat}) supports the excitation of small amplitude linear
waves:
\begin{equation*}
x_{n}(t)\sim \exp \left[ \mathrm{i}(\omega_{q}t-qn)\right] ,
\end{equation*}
with the wave number $q$ and the corresponding frequency spectrum $\omega
_{q}^{2}$ which, due to the underlying lattice, depends periodically on $q$:
\begin{equation*}
\omega _{q}^{2}=V^{\prime \prime }(0)+4W^{\prime \prime }(0)\sin ^{2}\left(
\frac{q}{2}\right) ,
\end{equation*}%
and its absolute value has always a finite upper bound. The maximum (Debye)
frequency of small amplitude waves is:
\begin{equation*}
\omega_q=\sqrt{V^{\prime \prime }(0)+4W^{\prime \prime }(0)}.
\end{equation*}

DBs exist for different types of potentials $W(x)$ and $V(x)$. DB solutions
are \cite{Flach03} (i) time-periodic: $\hat{x}_{n}(t+T_{b})=\hat{x}_{n}(t),$
and (ii) spatially localized: $\hat{x}_{|n|\rightarrow \infty }\rightarrow 0$%
. In addition, if the Hamiltonian $H$ is invariant under a finite
translation/rotation of any $x_{n}\rightarrow x_{n}+\lambda $, then \emph{%
discrete rotobreathers} may exist (see \cite{stmp96}), which are excitations
characterized by one or several sites in the breather center evolving in a
rotational state: $\hat{x}_{0}(t+T_{b})=\hat{x}_{0}(t)+\lambda $, while
outside this center the lattice is governed again by time periodic spatially
localized oscillations.

\hyphenation{cellular}
\section{Sine--Gordon solitons, kinks and breathers in living cellular structures}

In this section, we give the applications of the sine--Gordon equation (and the variety of its traveling--wave solutions), as spatiotemporal models of nonlinear excitations in living cellular structures.

\subsection{SGE--solitons in DNA}

In this subsection, we review the first three papers describing SGE--solitary
excitations in DNA.\footnote{The idea that it is possible that soliton excitations may suggest a discovery of a new mechanism in the duplication of DNA and the transcription
of messenger ribonucleic acid (mRNA) goes back to \cite{Englander}, who
demonstrated the existence of transiently open states in DNA and synthetic
polynucleotide double--helices, by hydrogen exchange measurements.} The
first two papers in this domain were published by \cite{Yomosa} and \cite%
{Zhang}\footnote{%
Note that, in the same period, Yakushevich \emph{et al.} performed their
SGE--solitary studies of DNA (see \cite{Yakushevich,YakushevichBk} and
references therein), focusing on the effects of weak inhomogeneities in
simple DNA fragments (consisting of uniform base sequences of a given type
followed by uniform base sequence of the other type), which were described
in terms of a parametrically--perturbed SGE.} -- incidentally, under the
same title, in the same journal (PRA), using two slightly-different
modifications of the same coupled SGE--system (\ref{2SGE}).

Firstly, Yomosa considered in \cite{Yomosa} (see also \cite{Yomosa2}) the
standard Watson--Crick double--helix $B-$form DNA model,\footnote{%
According to the Watson--Crick $B-$form DNA model, the two polynucleotide
strands forming a double helix are held together by hydrogen \textbf{H}%
-bonds. Yomosa was assuming that the \textbf{H}-bonding and the stacking
energies (consisting of the electrostatic, the exchange, the
charge-transfer, as well as the induction and dispersion interactions), were
roughly proportional to the overlaps of molecular orbitals.} in which
conformation and stability of DNA and the polynucleotide double helices are
determined by:\footnote{%
Here, the zero-level of the energies $E_{B}$ and $E_{S}$ are taken for the $%
B-$form of DNA and polynucleotide duplexes, while the mean energy of
distorted double and triple \textbf{H}-bonds in $A-T$ (adenine--thymine) and
$G-C$ (guanine--cytosine) base pairs is approximately represented (by the
formula for molecular association in liquids by \cite{Pople}) for the energy
of a distorted single \textbf{H}-bond.}

\begin{enumerate}
\item The energy $E_{B}$ of the hydrogen \textbf{H}-bonds between
inter-strand complementary base pairs, given by:%
\begin{equation*}
E_{B}=\sum_{n}B[1-\cos (\theta _{n}-\theta _{n}^{\prime }-\pi )],
\end{equation*}%
where $B$ is a parameter associated with the \textbf{H}-bond energy, while $%
\theta _{n}=\measuredangle (B_{n},P_{n})$ and $\theta _{n}^{\prime
}=\measuredangle (B_{n}^{\prime },P_{n}^{\prime })$ denote angles between
horizontal projections of the complementary base pairs and their
corresponding axes; \ and

\item The stacking energy $E_{S}$ between intra-strand adjacent bases, given
by:
\begin{equation*}
E_{S}=\sum_{n}S[1-\cos (\theta _{n}-\theta _{n-1}-\alpha _{0})]+S[1-\cos
(\theta _{n}^{\prime }-\theta _{n-1}^{\prime }-\alpha _{0})],
\end{equation*}%
where $S$ is a parameter associated with the stacking energy of DNA chains
and $\alpha _{0}=36^{\circ }.$
\end{enumerate}

Next, by adding the rotational kinetic energy:
\begin{equation*}
T_{\mathrm{rot}}=\frac{1}{2}\sum_{n}I[\dot{\theta}_{n}^{2}+\dot{\theta}%
_{n}^{\prime 2}],
\end{equation*}%
(where $I$ is the mean value of the moments of inertia $I_n$ of the bases
for the rotations around the axes $P$) to the potential energies $E_{B}$ and
$E_{S}$, the following SG--chain Hamiltonian for DNA and synthetic
polynucleotide double--helices was formulated:
\begin{eqnarray}
H &=&T_{\mathrm{rot}}+E_{B}+E_{S}=\frac{1}{2}\sum_{n}I[\dot{\theta}_{n}^{2}+%
\dot{\theta}_{n}^{\prime 2}]+\sum_{n}B[1-\cos (\theta _{n}-\theta
_{n}^{\prime }-\pi )]  \notag \\
&+&\sum_{n}\left\{ S[1-\cos (\theta _{n}-\theta _{n-1}-\alpha
_{0})]+S[1-\cos (\theta _{n}^{\prime }-\theta _{n-1}^{\prime }-\alpha
_{0})]\right\} .  \label{hamY}
\end{eqnarray}%
From the Hamiltonian (\ref{hamY}), via canonical Hamiltonian formalism, the
following two sets of coupled SGC--equations of motion were derived in \cite%
{Yomosa}:
\begin{eqnarray*}
I\ddot{\theta}_{n}+B\sin (\theta _{n}-\theta _{n}^{\prime }-\pi )+S[\sin
(\theta _{n}-\theta _{n-1}-\alpha _{0})-\sin (\theta _{n+1}-\theta
_{n}-\alpha _{0})] &=&0, \\
I\ddot{\theta}_{n}^{\prime }+B\sin (\theta _{n}-\theta _{n}^{\prime }-\pi
)+S[\sin (\theta _{n}^{\prime }-\theta _{n-1}^{\prime }-\alpha _{0})-\sin
(\theta _{n+1}^{\prime }-\theta _{n}^{\prime }-\alpha _{0})] &=&0.
\end{eqnarray*}%
Further, by linearizing this coupled ODE-system (assuming the smallness of
the angles $\theta _{n}-\theta _{n-1}-\alpha _{0}$ and $\theta _{n}^{\prime
}-\theta _{n-1}^{\prime }-\alpha _{0})$ and subsequently performing the
continuum limit:
\begin{equation}
\theta _{n}(t)\rightarrow \theta (x,t),\qquad \theta _{n}^{\prime
}(t)\rightarrow \theta ^{\prime }(x,t),  \label{cont}
\end{equation}%
Yomosa derived the following system of two coupled SGEs, of the
(slightly-modified) form of (\ref{2SGE}):%
\begin{eqnarray}
I\theta _{tt}-S\theta _{xx} &=&-B\sin (\theta -\theta ^{\prime }-\pi ),
\label{SGE2Y} \\
I\theta _{tt}^{\prime }-S\theta _{xx}^{\prime } &=&B\sin (\theta -\theta
^{\prime }-\pi ).  \notag
\end{eqnarray}%
Unfortunately, in his time, Yomosa was not able solve the coupled system (%
\ref{SGE2Y}), so he took the difference of the two SGEs and obtained the
following single SGE representing a dynamic (plane) base-rotator model:\
\begin{equation}
\frac{1}{v_{0}^{2}}\phi _{tt}=\phi _{xx}-\frac{1}{l^{2}}\sin \phi ,\qquad
\text{where \ (}v_{0}=\sqrt{\frac{S}{I}},\text{ \ }l=\sqrt{\frac{S}{2B}}).
\label{brm}
\end{equation}%
Finally, by imposing the following boundary condition:
\begin{eqnarray*}
\cos \phi &=&1\qquad \text{\textrm{for} \ }(\phi =2\pi n,\text{ \ \ }n=0,\pm
1,...)\qquad \\
\text{\textrm{at} \ }\xi &=&\pm \infty \qquad (\text{\textrm{at} \ }x=\pm
\infty \quad \text{\textrm{for all \ }}t),
\end{eqnarray*}%
the following traveling solitary-wave solutions of (\ref{brm}) were obtained
in the form (\ref{tabor}) of a\emph{kink--antikink} pair:
\begin{equation*}
\phi (x,t)=4\,\arctan \exp \left[ \pm \frac{\left( \xi -\xi _{0}\right) }{%
\sqrt{1-v/v_{0}^{2}}l}\right] .\text{ \ }
\end{equation*}

Secondly, Zhang clarified in \cite{Zhang} the pioneering (and therefore
somewhat-messy) approach of Yomosa and proposed the following modified
SGC--Hamiltonian for the $B-$form DNA double--helix (rewritten here in above
Yomosa's notation for consistency):%
\begin{eqnarray}
H &=&\frac{1}{2}\sum_{n}I[\dot{\theta}_{n}^{2}+\dot{\theta}_{n}^{\prime
2}]+\sum_{n}V(\theta _{n},\theta _{n}^{\prime })  \label{hamZ} \\
&+&\frac{1}{2}\sum_{n}\left[ S(\theta _{n}-\theta _{n-1})^{2}+S(\theta
_{n}^{\prime }-\theta _{n-1}^{\prime })^{2}\right] ,  \notag
\end{eqnarray}%
where $V(\theta _{n},\theta _{n}^{\prime })$ is the inter-strand interaction
energy in $n$th base pair, given by:\footnote{%
Here the zero-level of the energy is taken for the $B-$form DNA, the same
way as Yomosa did.}
\begin{eqnarray*}
V(\theta _{n},\theta _{n}^{\prime }) &=&B[1-\cos (\theta _{n}-\theta
_{n}^{\prime })]+\lambda (1-\cos \theta _{n})+\lambda (1-\cos \theta
_{n}^{\prime }) \\
&+&\beta \left\{ 3(1-\cos \theta _{n}\cos \theta _{n}^{\prime })-[1-\cos
(\theta _{n}-\theta _{n}^{\prime })]\right\} .
\end{eqnarray*}%
From the Hamiltonian (\ref{hamZ}), the following SGC--equations of motion
were derived in \cite{Zhang}:%
\begin{eqnarray}
&&I\ddot{\theta}_{n}+B\sin (\theta _{n}-\theta _{n}^{\prime })+\beta \lbrack
3\sin \theta _{n}\cos \theta _{n}^{\prime }-\sin (\theta _{n}-\theta
_{n}^{\prime })]+\lambda \sin \theta _{n}  \notag \\
&&\qquad =S(\theta _{n+1}-2\theta _{n}+\theta _{n-1}),  \label{sgcpert} \\
&&I\ddot{\theta}_{n}^{\prime }-B\sin (\theta _{n}-\theta _{n}^{\prime
})+\beta \lbrack 3\cos \theta _{n}\sin \theta _{n}^{\prime }+\sin (\theta
_{n}-\theta _{n}^{\prime })]+\lambda \sin \theta _{n}^{\prime }  \notag \\
&&\qquad =S(\theta _{n+1}^{\prime }-2\theta _{n}^{\prime }+\theta
_{n-1}^{\prime }).  \notag
\end{eqnarray}%
By performing the approximation (\ref{cont}), Zhang introduced the continuum
variables: $\theta \ $and $\theta ^{\prime }$. Subsequently, by introducing
new variables: $\phi =\theta +\theta ^{\prime },$ $\psi =\theta -\theta
^{\prime },$ he obtained the following system of two perturbed and coupled
SGEs:
\begin{eqnarray*}
\phi _{xx}-(1/c_{0}^{2})\phi _{tt} &=&(1/l^{2})\sin \phi +(2/d^{2})\sin
(\phi /2)\cos (\psi /2), \\
\psi _{xx}-(1/c_{0}^{2})\psi _{tt} &=&(1/l^{2})\sin \psi +(2/d^{2})\sin
(\psi /2)\cos (\phi /2), \\
\text{\textrm{where}}\quad c_{0} &=&\sqrt{S/I},\quad l=\sqrt{S/(3\beta )},\
\quad d=\sqrt{S/\lambda },
\end{eqnarray*}%
which cannot be solved analytically. So, by setting $\lambda =0$ in (\ref%
{sgcpert}), Zhang arrived at the following system of two independent SGEs
[(with $Q=(2B+\beta )/(3\beta )$]:
\begin{eqnarray}
\phi _{xx}-(1/c_{0}^{2})\phi _{tt} &=&(1/l^{2})\sin \phi ,  \label{zh} \\
\psi _{xx}-(1/c_{0}^{2})\psi _{tt} &=&(Q/l^{2})\sin \psi ,  \notag
\end{eqnarray}%
with the simple solution of a single soliton with velocity $c$:
\begin{eqnarray}
&&\phi _{0}^{\pm }(x,t)=4\,\arctan \exp \left( \pm z\right) ,\qquad \psi
_{0}^{\pm }(x,t)=4\,\arctan \exp \left( \pm \sqrt{Q}\,z\right) ,
\label{zsol} \\
&&\text{\textrm{where}}\quad z=(\gamma /l)(\xi -\xi _{0}),\quad \xi
=x-ct,\quad \gamma =1/\sqrt{1-c^{2}/c_{0}^{2}}.  \notag
\end{eqnarray}%
Finally, by returning to original continuum variables $\theta \ $and $\theta
^{\prime },$ from (\ref{zsol}), Zhang obtained a set of $\pi $-kink/antikink
and $2\pi $-kink/antikink solutions (see \cite{Zhang} for more technical
details).

The third (and most-cited) paper in this domain (of SGE--solitary
excitations in DNA) was \cite{Salerno91} (see also \cite{Salerno92,Salerno94}%
), who proposed a discrete SGC model for DNA-promoter dynamics. Salerno
introduced the following SGC--Hamiltonian (slightly refined from Yomosa's
and Zhang's Hamiltonians):%
\begin{eqnarray}
H &=&\frac{1}{2}\sum_{n=1}^{N}I[\dot{\psi}_{n}^{2}+\dot{\theta}%
_{n}^{2}]+\sum_{n=1}^{N}K\left[ (\psi _{n+1}-\psi _{n})^{2}+(\theta
_{n+1}-\theta _{n})^{2}\right]  \notag \\
&+&\sum_{n=1}^{N}\eta _{n}[1-\cos (\psi _{n}-\theta _{n})],  \label{hamS}
\end{eqnarray}%
where $N$ is the number of base-pairs in the SGC and $K$ is the backbone
spring constant along both DNA--helices. $\eta _{n}$ is a nonlinear
parameter used to model the strength of \textbf{H}-bonds between
complementary bases, chosen according to the following rule: $\eta
_{n}=\lambda _{n}\beta $ with $\lambda _{n}=2$ if it refers to $A-T$ or $T-A$
pairs, $\lambda _{n}=3$ otherwise, with $\beta $ a free parameter. From the
Hamiltonian (\ref{hamS}), the following SGC--equations of motion were
derived:
\begin{eqnarray}
I\ddot{\psi}_{n} &=&K(\psi _{n+1}-2\psi _{n}+\psi _{n-1})-\frac{\beta }{2}%
\lambda _{n}\sin (\psi _{n}-\theta _{n}),  \label{sgcS} \\
I\ddot{\theta}_{n} &=&K(\theta _{n+1}-2\theta _{n}+\theta _{n-1})-\frac{%
\beta }{2}\lambda _{n}\sin (\theta _{n}-\psi _{n}).  \notag
\end{eqnarray}%
Further, from (\ref{sgcS}), the following SGC--equation of motion is
obtained for\ the angle difference: $\phi _{n}=\psi _{n}-\theta _{n}$,
between complementary bases:%
\begin{equation}
\ddot{\phi}_{n}=\phi _{n+1}-2\phi _{n}+\phi _{n-1}-\frac{\beta }{K}\lambda
_{n}\sin \phi _{n}.  \label{odeS}
\end{equation}%
We remark that the ODE (\ref{odeS}) has the standard form of (\ref{SGC});
from it, in the continuum limit, the standard SGE (\ref{SGE1}) is obtained,
with exact soliton solutions (as described in the subsection \ref{kinks}
before). Salerno used the ODE--model (\ref{odeS}) to study nonlinear wave
dynamics of the $T7A1-$DNA promoter (see \cite{Salerno91} for further
technical details).

\subsection{SGE--solitons in protein folding}

For over a decade, it has been known that nonlinear excitations can
influence conformational dynamics of biopolymers; e.g., the effective
bending rigidity of a biopolymer chain could lead to a buckling instability
\cite{mingaleev}. Subsequently, several models have been proposed to explain
such protein transition (see, e.g. \cite{sulaiman2} and references therein).

In this subsection, we review two protein--folding dynamics papers.
Firstly, it was suggested in \cite{Caspi99} that protein folding may be
mediated via interaction of the protein (molecular) chain with SGE--solitons
which propagate along the chain. Local potential energy of the chain is
modeled by an asymmetric double-well potential:%
\begin{equation*}
V(\varphi )=\varepsilon (\varphi +\delta )^{2}(\varphi ^{2}-{\frac{2}{3}}%
\varphi \delta +{\frac{1}{3}}\varphi ^{2}-2),
\end{equation*}%
where the scalar variable $\varphi $ represents local conformation of the
protein, $\varepsilon $ is a small positive parameter, while $\delta $ is
the asymmetry parameter (ranging from $-1$ to $1)$. The two minima of the
potential, corresponding to the local $\alpha $- and $\beta $-conformations
of the chain, are positioned at $\varphi =\pm 1$ and the energy-difference
between them is: \ $\Delta E={\frac{16}{3}}\varepsilon \delta .$

Solitonic excitations are realized in \cite{Caspi99} by an additional,
dissipative SGE--field $\phi (x),$ where $x$ is the position along the
protein. The following interaction potential (with the positive parameter $%
\Lambda $) is used to mediate interaction between the two fields:%
\begin{equation*}
u(\phi ,\varphi )={\frac{\Lambda }{\Lambda +1}}(1-\cos \phi )\varphi ^{2}.
\end{equation*}%
The following dissipative equations of motion are derived \cite{Caspi99}:
\begin{eqnarray}
\phi _{tt} &=&\phi _{xx}-{\frac{1+\Lambda \varphi ^{2}}{1+\Lambda }}\sin
\phi -\gamma _{\phi }\phi _{t},  \label{eq_of_motion} \\
m\varphi _{tt} &=&-4\varepsilon (\varphi +\delta )(\varphi ^{2}-1)-{\frac{%
2\Lambda }{1+\Lambda }}\varphi (1-\cos \phi )-\gamma _{\varphi }\varphi _{t},
\notag
\end{eqnarray}%
where $\gamma _{\phi }\phi _{t}$ and $\gamma _{\varphi }\varphi _{t}$ are
dissipative terms. In the small interaction limit (ignoring $\gamma _{\phi
}\phi _{t}$), it is chosen:%
\begin{equation*}
\phi (x,t)=f(x-vt)+\Delta \theta (x,t),\qquad \varphi (x,t)=1+\Delta \varphi
(x,t),
\end{equation*}%
where $f(z)=4\arctan (\mathrm{e}^{-z})$ is the usual SGE--kink, moving with
velocity $v$. For $\varphi (x,t),$ the following approximate sech--soliton
solution is obtained:
\begin{equation*}
\Delta \varphi \simeq -{\frac{4\Lambda /m}{v^{2}+(\omega /2)^{2}}}\,{\frac{1%
}{\cosh ^{2}(x-vt)}};
\end{equation*}%
so, near the center of the $\phi $ kink, $\varphi $ is pushed away from its
local minima $\varphi =1$ towards the other local minima. A localized static
solution of (\ref{eq_of_motion})\ is found to be:
\begin{equation*}
\phi =4\tan ^{-1}{\frac{1}{q+\sqrt{1+q^{2}}}},\qquad \varphi ^{2}=1-{\frac{%
\Lambda }{\varepsilon (1+\Lambda )}}\,{\frac{1}{1+q^{2}}},
\end{equation*}%
where $q\equiv \sqrt{1-a}\,\sinh (x-x_{0})$, $a=\Lambda ^{2}/[2\varepsilon
(1+\Lambda )^{2}]{;}$ for more technical details, see \cite{Caspi99}.\\

More recently, a Lagrangian field--theory based modeling approach to protein folding
has been proposed in \cite{Januar12}. They proposed the protein Lagrangian
including three terms:

(i) nonlinear unfolding $\phi ^{4}-$protein at the initial state:%
\begin{equation*}
\mathcal{L}_{\mathrm{unf}}=\frac{1}{2}\left( \partial _{\mu }\phi \right)
^{\dagger }\left( \partial ^{\mu }\phi \right) +\frac{m_{\phi }^{4}}{\lambda
_{\phi }}\left[ 1-\cos \left( \frac{\sqrt{\lambda _{\phi }}}{m_{\phi }}|\phi
|\right) \right] ;
\end{equation*}

(ii) nonlinear sources injected into the backbone, modeled by $\psi ^{4}$
self-interaction:%
\begin{equation*}
\mathcal{L}_{\mathrm{src}}=\frac{1}{2}\left( \partial _{\mu }\psi \right)
^{\dagger }\left( \partial ^{\mu }\psi \right) +\frac{\lambda _{\psi }}{4!}%
\,(\psi ^{\dagger }\psi )^{2};
\end{equation*}

(iii) the interaction term (with the coupling constant $\Lambda$):%
\begin{equation*}
\mathcal{L}_{\mathrm{int}}=-\Lambda \,(\phi ^{\dagger }\phi )(\psi ^{\dagger
}\psi ).
\end{equation*}
The total potential (from all three terms) reads:%
\begin{equation*}
V_{\mathrm{tot}}(\phi ,\psi )=\frac{m_{\phi }^{4}}{\lambda _{\phi }}\left[
1-\cos \left( \frac{\sqrt{\lambda _{\phi }}}{m_{\phi }}|\phi |\right) \right]
+\frac{\lambda _{\psi }}{4!}\,(\psi ^{\dagger }\psi )^{2}-\Lambda \,(\phi
^{\dagger }\phi )(\psi ^{\dagger }\psi ).
\end{equation*}%
Assuming that $\lambda _{\phi }$is small enough to be approximately at the
same order with $\lambda _{\psi }$, the first term can be expanded in term
of $\sqrt{\lambda _{\psi }}$, giving (up to the 2$^{nd}$ order accuracy):%
\begin{equation*}
V_{\mathrm{tot}}(\phi ,\psi )\approx \frac{m_{\phi }^{2}}{2}\,\phi ^{\dagger
}\phi -\frac{\lambda _{\phi }}{4!}\,(\phi ^{\dagger }\phi )^{2}+\frac{%
\lambda _{\psi }}{4!}\,(\psi ^{\dagger }\psi )^{2}-\Lambda \,(\phi ^{\dagger
}\phi )(\psi ^{\dagger }\psi ),
\end{equation*}%
from which the total Lagrangian: $\mathcal{L}_{\mathrm{tot}}=\mathcal{L}_{%
\mathrm{unf}}+\mathcal{L}_{\mathrm{src}}+\mathcal{L}_{\mathrm{int}}$ can be
(up to the 2$^{nd}$-order accuracy)\ approximated by:\qquad
\begin{eqnarray}
\mathcal{L}_{\mathrm{tot}}(\phi ,\psi ) &=&\frac{1}{2}\left[ \left( \partial
_{\mu }\phi \right) ^{\dagger }\left( \partial ^{\mu }\phi \right) +\left(
\partial _{\mu }\psi \right) ^{\dagger }\left( \partial ^{\mu }\psi \right) %
\right]   \label{Ltot} \\
&&+\frac{m_{\phi }^{2}}{2}\,\phi ^{\dagger }\phi -\frac{\lambda _{\phi }}{4!}%
\,(\phi ^{\dagger }\phi )^{2}+\frac{\lambda _{\psi }}{4!}\,(\psi ^{\dagger
}\psi )^{2}-\Lambda \,(\phi ^{\dagger }\phi )(\psi ^{\dagger }\psi ).  \notag
\end{eqnarray}%
From the Euler--Lagrangian PDEs for the total Lagrangian (\ref{Ltot}):%
\begin{equation*}
\frac{\partial \mathcal{L}_{\mathrm{tot}}}{\partial |\phi |}-\partial _{\mu }%
\frac{\partial \mathcal{L}_{\mathrm{tot}}}{\partial (|\partial _{\mu }\phi |)%
}=0,\;\;\;\;\frac{\partial \mathcal{L}_{\mathrm{tot}}}{\partial |\psi |}%
-\partial _{\mu }\frac{\partial \mathcal{L}_{\mathrm{tot}}}{\partial
(|\partial _{\mu }\psi |)}=0,
\end{equation*}%
the following coupled and perturbed SGE and (nonlinear) KGE with cubic
forcing are derived:%
\begin{eqnarray}
\phi _{tt} &=&\phi _{xx}-\frac{m_{\phi }^{3}}{\sqrt{\lambda _{\phi }}}\,\sin
\left( \frac{\sqrt{\lambda _{\phi }}}{m_{\phi }}|\phi |\right) +2\Lambda
\,|\phi ||\psi |^{2},  \label{eq:eomc} \\
\psi _{tt} &=&\psi _{xx}-\frac{\lambda _{\psi }}{6}\,|\psi |^{3}+2\Lambda
\,|\psi ||\phi |^{2},  \label{eq:eoms}
\end{eqnarray}%
where $\lambda _{\phi }-$ and $\lambda _{\psi }-$terms determine
nonlinearities of backbone and source, respectively.

Numerical solution of the two coupled (1+1) PDEs, (\ref{eq:eomc})--(\ref%
{eq:eoms}), with the following boundary conditions:
\begin{equation}
\begin{array}{lcl}
\phi (0,t)=\phi (L,t)=0\;\;\text{and}\;\;\psi (0,t)=\psi (L,t)=0 & \text{for}
& 0\leq t\leq b, \\
\phi (x,0)=p(x)\;\;\text{and}\;\;\psi (x,0)=f(x) & \text{for} & 0\leq x\leq
L, \\
\phi _{t}(x,0)=q(x)\;\;\text{and}\;\;\psi _{t}(x,0)=g(x) & \text{for} &
0<x<L,%
\end{array}
\label{eq:bc}
\end{equation}%
(where $p(x)$, $q(x)$, $f(x)$ and $g(x)$ are auxiliary functions), would
describe the contour of conformational changes for protein folding. It was
performed in \cite{Januar12} using the following forward finite differences:%
\begin{eqnarray*}
\phi _{i,j+1} &=&2\phi _{i,j}-\phi _{i,j-1}+\Delta t^{2}\left( \frac{\phi
_{i+1,j}-2\phi _{i,j}+\phi _{i-1,j}}{\Delta x^{2}}+2\Lambda \psi
_{i,j}^{2}\phi _{i,j}\right.  \\
&&\left. -\frac{m_{\phi }^{3}}{\hbar ^{3}\sqrt{\lambda _{\phi }}}\sin \left(
\frac{\sqrt{\lambda _{\phi }}}{m_{\phi }}\,\phi _{i,j}\right) \right) , \\
\psi _{i,j+1} &=&2\psi _{i,j}-\psi _{i,j-1}+\Delta t^{2}\left( \frac{\psi
_{i+1,j}-2\psi _{i,j}+\psi _{i-1,j}}{\Delta x^{2}}+2\Lambda w_{i,j}^{2}\psi
_{i,j} -\frac{\lambda _{\psi }}{6}\psi _{i,j}^{3}\right) ,
\end{eqnarray*}%
where the values at the first time-step $t_{1}$ are given by the boundary
conditions (\ref{eq:bc}), while the values at the second time-step $t_{2}$
are determined by the 2$^{nd}$-order Taylor expansion:
\begin{eqnarray*}
\phi _{i,2} &=&p_{i}-\Delta tq_{i}+\frac{1}{2}\Delta t^{2}\left( \frac{%
p_{i+1}-2p_{i}+p_{i-1}}{\Delta x^{2}}+2\Lambda f_{i}^{2}p_{i}\right.  \\
&&\left. -\,\frac{m_{\phi }^{3}}{\hbar ^{3}\sqrt{\lambda _{\phi }}}\sin
\left( \frac{\sqrt{\lambda _{\phi }}}{m_{\phi }}\,p_{i}\right) \right)
,\qquad (\mathrm{for\ }i=2,3,\cdots ,N-1), \\
\psi _{i,2} &=&f_{i}-\Delta tg_{i}+\frac{1}{2}\Delta t^{2}\left( \frac{%
f_{i+1}-2f_{i}+f_{i-1}}{\Delta x^{2}}+2\Lambda p_{i}^{2}f_{i}-\frac{\lambda
_{\psi }}{6}f_{i}^{3}\right) .
\end{eqnarray*}%
For more technical details, see \cite{Januar12}.

\subsection{SGE--solitons in microtubules}

In this subsection, we review two most-significant papers describing
SGE--solitary excitations in microtubules (MTs),\footnote{%
MTs are major cytoskeletal proteins assembled from the tubulin protein that
plays a crucial role in all eukaryotic cells. MT functions include cellular
orientation, structure and guidance of membrane and cytoplasmic movement,
which are crucial effects to cell division, morphogenesis, and embryo
development. MT structure is a hollow cylinder that typically involves 13
protofilaments (of protein dimers called tubulins.). Each protofilament is
formed from tubulin molecules arranged in a 'head-to-tail joint' fashion.
The inner and the outer diameters of the cylinder are 15nm and 25nm, while
its length may span dimensions from the order of micrometer to the order of
millimeter. Each dimer is an electric dipole whose length and longitudinal
component of the electric dipole moment are 8nm and 337Debye, respectively.
The constituent parts of the dimers are $\alpha $ and $\beta $ tubulins,
corresponding to positively and negatively charged sides, respectively (see
\cite{Dustin}, as well as references in \cite%
{Sataric93,Sataric98,Sataric03,Sataric12}).} and then point-out to some
related quantum studies of neural MTs.

\subsubsection{Soliton dynamics in MTs}

To the best of our knowledge, the first paper describing soliton dynamics in
MTs was \cite{Sataric93} (see also \cite{Sataric98,Sataric03,Sataric12}; for a recent review, see \cite{QNNbk}), in
which Satari\'{c} \emph{et al.} developed a ferroelectric model of neural
MTs where the motion of MT subunits is reduced to a single longitudinal DOF
per dimer at a position $n$, denoted by $\phi _{n}.$ The overall effect of
the surrounding dimers on a dipole at a chosen site $n$ can be qualitatively
described by the following double-well quartic potential:
\begin{equation}
V(\phi _{n})=-\frac{1}{2}A\phi _{n}^{2}+\frac{1}{4}B\phi _{n}^{4}\,,
\label{dbwell}
\end{equation}%
where $A$ and $B$ are real parameters ($A$ is a linear function of
temperature). As an electrical dipole, a dimer in the intrinsic electric
field of the MT acquires the additional potential energy given by:
\begin{equation*}
V_{\mathrm{el}}=-qE\phi _{n}\,,
\end{equation*}%
where $q=18\times 2e$ ($e=$ electron charge) denotes the effective mobile
charge of a single dimer and $E$ is the magnitude of the intrinsic electric
field. In addition, two more MT--related energies can also be defined:%
\newline
(i) the potential energy of restoring strain--forces between adjacent dimers
in the protofilament (with a unique spring/stiffness constant $k$):
\begin{equation*}
V_{\mathrm{str}}=\frac{1}{2}k(\phi _{n+1}-\phi _{n})^{2},\qquad \text{%
\textrm{and}}
\end{equation*}%
(ii) the (velocity $\dot{\phi}_{n}-$dependent) kinetic energy associated
with the longitudinal displacements of constituent dimers with unique mass $%
m $:
\begin{equation*}
T(\dot{\phi}_{n})=\frac{1}{2}\sum_{n=1}^{N}m\dot{\phi}_{n}^{2},
\end{equation*}%
where $N$ is the total number of constituent dimers in the \emph{%
microtubular chain} (MTC). The full MTC--Hamiltonian is now given by:%
\begin{eqnarray}
H &=&T(\dot{\phi}_{n})+V_{\mathrm{str}}+V(\phi _{n})+V_{\mathrm{el}}
\label{hamSat} \\
&=&\sum_{n=1}^{N}\left[ \frac{1}{2}m\dot{\phi}_{n}^{2}+\frac{1}{2}k(\phi
_{n+1}-\phi _{n})^{2}-\frac{1}{2}A\phi _{n}^{2}+\frac{1}{4}B\phi
_{n}^{4}\,-qE\phi _{n}\right] .  \notag
\end{eqnarray}%
Also, in order to derive a realistic equation of motion for the MTC, it is
indispensable to include the viscous force: $F_{\mathrm{v}}(\dot{\phi}%
_{n})=-\gamma \dot{\phi}_{n}$, with the damping coefficient $\gamma .$%
\footnote{%
We remark that the friction $\gamma $ should be viewed as an environmental
effect, which does not lead to energy dissipation; this is a well-known
result, relevant to energy transfer in biological systems \cite{lal}.}

Using the Hamiltonian (\ref{hamSat}) together with the damping force $F_{%
\mathrm{v}}(\dot{\phi}_{n}),$ and subsequently performing the continuum
limit (with equilibrium spacing $r_{0}$ between adjacent dimers):
\begin{eqnarray*}
\phi _{n}(t) &\rightarrow &\phi (x,t), \\
\phi _{n+1}(t) &\rightarrow &\phi (x,t)+r_{0}\phi _{x}(x,t)+\frac{1}{2}%
r_{0}^{2}\phi _{xx}(x,t)+...,
\end{eqnarray*}%
-- Satari\'{c} \emph{et al.} finally derived their nonlinear (forced and
damped) wave equation: \
\begin{equation}
m\phi _{tt}+\gamma \phi _{t}-kr_{0}^{2}\phi _{xx}=A\phi -B\phi ^{3}+qE.
\label{wavSat}
\end{equation}%
The importance of the electric force term $qE$ lies in the fact that the PDE
(\ref{wavSat}) admits soliton solutions with no energy loss, which acquires
the form of a traveling wave, and can be expressed by defining a normalized
displacement field \cite{lal}:
\begin{equation*}
\psi (\xi )=\frac{\phi (\xi )}{\sqrt{A/B}},\quad \text{\textrm{with\ \ }}%
\left( \xi =\alpha (x-vt),\,\alpha =\sqrt{\frac{|A|}{m(v_{0}^{2}-v^{2})}}%
\right) ,
\end{equation*}%
where $v_{0}=\sqrt{k/m}r_{0}$ is the sound velocity and $v$ is the
soliton--propagation velocity. In terms of the $\psi (\xi )$ variable, the
wave equation (\ref{wavSat}) reduces to a damped anharmonic oscillator ODE:
\begin{eqnarray}
\psi ^{\prime \prime } &+&\rho \psi ^{\prime 3}+\psi +\sigma =0,\qquad \text{%
\textrm{where}}  \label{ten} \\
\rho &=&\gamma v[m|A|(v_{0}^{2}-v^{2})]^{-\frac{1}{2}},\qquad \sigma =q\sqrt{%
B}|A|^{-3/2}E,  \notag
\end{eqnarray}%
which has a unique bounded solution \cite{Sataric93}:%
\begin{eqnarray}
&&\psi (\xi )=a+\frac{b-a}{1+\exp \left( \frac{b-a}{\sqrt{2}}\xi \right) }%
,\qquad \text{\textrm{such that:}}  \label{eleven} \\
&&(\psi -a)(\psi -b)(\psi -d)=\psi ^{3}-\psi -\left( \frac{q\sqrt{B}}{%
|A|^{3/2}}E\right) .  \label{par}
\end{eqnarray}%
So the kink propagates along the protofilament axis with fixed velocity: \
\begin{equation*}
v=v_{0}/\sqrt{1+2\gamma ^{2}/(9d^{2}m|A|)},
\end{equation*}
which depends on the strength of the electric field $E$ via (\ref{par}). The
total \emph{conserved energy} of the kink (\ref{eleven}) is given by:%
\footnote{%
The first term of (\ref{energy}) expresses the binding energy of the kink
and the second the resonant transfer energy (not that in realistic
biological models, the sum of these two terms dominate over the third term,
being of order of $1eV$).}
\begin{eqnarray}
E &=&\frac{2\sqrt{2}}{3}\frac{A^{2}}{B}+\frac{\sqrt{2}}{3}k\frac{A}{B}+\frac{%
1}{2}m^{\ast }v^{2},\qquad \text{\textrm{where}}  \label{energy} \\
m^{\ast } &=&\frac{4}{3\sqrt{2}}\frac{mA\alpha }{r_{0}B}\qquad \text{is kink}%
^{\prime }\text{s \emph{effective} mass.}  \notag
\end{eqnarray}
For the further development of the theory, with kink-antikink waves
traveling in opposite directions along the MTC, see \cite%
{Sataric93,MN97,Sataric98,Sataric03,Sataric12}.

Now, from our general SGE perspective, Satari\'{c} model (\ref{wavSat}) can
be approximated by the perturbed SGE (\ref{sg2}), rewritten here for our
readers' convenience:
\begin{equation*}
\phi _{tt}+\gamma \phi _{t}-\phi _{xx}+\sin \phi =F,
\end{equation*}%
-- if we apply the following assumptions:\newline
(i) normalized units: $m=k=r_{0}=1$;\newline
(ii) electrical force: $qE=F\equiv F(x,t);$ ~and\newline
(iii) using the (first two terms of the) Taylor--series expansion of the
sine term:
\begin{equation}
\sin \phi =\phi -\frac{1}{6}\phi ^{3}+O(\phi ^{4})\approx A\phi -B\phi ^{3}.
\label{sinTaylor}
\end{equation}%
Using assumptions (i)--(ii) and approximation (\ref{sinTaylor}), all results
of the section \ref{pertSG} (illustrated with the simulation Figures \ref%
{sgMma2}--\ref{sgMma6}) are ready to be employed for the further SGE
analysis of solitary excitations in microtubules.

The second paper describing soliton dynamics in MTs was \cite{Chou94}. Under
similar assumptions as \cite{Sataric93}, they proposed the same expresions
for the kinetic energy $T(\dot{\phi}_{n})$\ and potential energy $V_{\mathrm{%
str}}$ of restoring strain--forces between adjacent dimers. However, in
contrast to the quartic potential (\ref{dbwell}), Chou \emph{et al.}
followed the recipes from solid state physics \cite{Dodd} and expressed the
interaction for the $n$th tubulin molecule of a protofilament by the
following periodic (effective) potential:%
\begin{equation*}
V(\phi _{n})=V_{0}\left[ 1-\cos \left( \frac{2\pi \phi _{n}}{a_{0}}\right) %
\right] \,,
\end{equation*}%
where $V_{0}$ is the half-height of the potential energy barrier, $\phi _{n}$
is the displacement of the $n$th tubulin molecule within a particular
protofilament and $a_{0}=8$nm is the distance between the centers of two
neighboring tubulin molecules along a protofilament. In this way, they
defined the following MTC--Hamiltonian:%
\begin{eqnarray}
H &=&T(\dot{\phi}_{n})+V_{\mathrm{str}}+V(\phi _{n})  \label{hamC} \\
&=&\sum_{n=1}^{N}\left[ \frac{1}{2}m\dot{\phi}_{n}^{2}+\frac{1}{2}k(\phi
_{n+1}-\phi _{n})^{2}+V_{0}\left( 1-\cos \frac{2\pi \phi _{n}}{a_{0}}\right) %
\right] .  \notag
\end{eqnarray}%
Using canonical Hamilton's equations, from (\ref{hamC}) they derived the
MTC--equations of motion for the $n$th tubulin molecule within a
protofilament (for $n=1,2,...N$):%
\begin{equation}
m\ddot{\phi}_{n}=k(\phi _{n+1}-2\phi _{n}+\phi _{n-1})-\frac{2\pi V_{0}}{%
a_{0}}\sin \frac{2\pi \phi _{n}}{a_{0}}.  \label{mtcCh}
\end{equation}%
In the continuum limit: $\phi _{n}(t)\rightarrow \phi (x,t),$ the
MTC--equations of motion (\ref{mtcCh}) reduce to the SGE:%
\begin{eqnarray}
m\phi _{tt} &=&ka_{0}^{2}\phi _{xx}-\left( \frac{2\pi }{a_{0}}\right)
^{2}V_{0}\sin \phi ,\qquad \text{\textrm{or}}  \label{sgCh} \\
\phi _{tt} &=&\frac{1}{c^{2}}\phi _{xx}-\frac{1}{l^{2}}\sin \phi ,\qquad
\text{\textrm{with} \ }\left( c^{2}=\frac{ka_{0}^{2}}{m},~~\frac{1}{l^{2}}=%
\frac{4\pi ^{2}V_{0}}{ka_{0}^{4}}\right) ,  \notag
\end{eqnarray}%
which is the same SGE as (\ref{zh}) that was used by \cite{Zhang} for
DNA--solitons, with kink--antikink solutions (\ref{zsol}). Using (\ref{sgCh}%
), Chou \emph{et al.} showed that there was a very high and narrow peak at
the center of the kink width, implying that a tubulin molecule would have
its maximum momentum when it reaches the top of the periodic potential (for
more technical details, see \cite{Chou94}).

In particular, in the case of \emph{neural MTs}, possibility for
sub-neuronal processing of information by cytoskeletal \emph{tubulin tails}
has been proposed by \cite{GPG07}, by showing that local electromagnetic
field supports information that could be converted into specific protein
tubulin-tail conformational states. Long-range collective coherent behavior
of the tubulin tails could be modeled in the form of sine-Gordon kinks,
antikinks or breathers that propagate along the microtubule outer surface,
and the tubulin-tail soliton collisions could serve as elementary
computational gates that control cytoskeletal processes. The authors of \cite%
{GPG07} have used the results of \cite{Abdalla}, combined with\ the \emph{%
elastic ribbon SGE--model} of \cite{Dodd}. Applying B\"{a}cklund
transformations (\ref{bt}) they found 2- and 3-soliton solutions, as well as
their elastic collisions. They developed $Maple^{TM}$-based animations of a
whole `zoo' of colliding solitons, including kink/antikink pairs and three
types of breathers: (i) a standing breather, (ii) a traveling large
amplitude breather and (iii) a traveling small amplitude breather.\footnote{%
The standing breather soliton could be obtained \emph{in vivo} in
experiments in which the electric-field vector acts perpendicular to the
microtubule $z-$axis. If a local vortex of the electromagnetic field is
created somewhere in the neuron, then the exerted action of the electric
field vector along the z-axis will be zero and no traveling soliton would be
born. This standing breather, swinging at certain tubulin tail could
catalyze attachment/detachment of microtubule associated proteins and
promote or inhibit the kinesin walk.}

\subsubsection{Liouville's stringy time--arrow in neural MTs}

Now, recall that the term cubic in $\psi $ in the equation of motion (\ref%
{ten}) was responsible for the appearance of a kink-like classical solution.
Let us formulate a Liouville $(1+1)-$string theory of the neural
MT--complex, following \cite{MN97}, and consider a general polynomial in $T$
equation of motion for a static tachyon:
\begin{equation}
T^{\prime \prime }(\xi )+\rho T^{\prime }(\xi )=P(T),  \label{polyn}
\end{equation}%
(where $\xi $ is some space-like coordinate and $P(T)$ is a polynomial of
degree $n)$, in which the `friction' term $T^{\prime }$ expresses a
Liouville derivative. In this interpretation of the Liouville field as a
local scale on the world-sheet it is natural to assume that the
single-derivative term expresses the non-critical string $\beta $-function,
and hence is itself a polynomial $R$ of degree $m$: $T^{\prime }(\xi )=R(T)$%
. Such equations lead to kink-like solutions \cite{curiouseq}:
\begin{eqnarray}
T(\xi ) &=&\frac{1}{2a_{4}}\{\mathrm{sgn}(a_{2}a_{4})a_{2}\mathrm{tanh}[%
\frac{1}{2}a_{2}(x-vt)]-a_{2}\},  \label{kink} \\
\text{\textrm{where \ }}v &=&\frac{A_{3}-3a_{2}a_{4}}{a_{4}}\text{ \ \textrm{%
is the velocity,}}  \notag
\end{eqnarray}%
which is a universal behavior for biological systems \cite{lal}, showing the
existence of a scheme which admits kink--like solutions for energy transfer
without dissipation in cells. The structure of the equation (\ref{polyn}),
which leads to kink-like solutions (\ref{kink}), is generic for Liouville
strings in non-trivial background space-times.

According to the conventional \emph{Liouville theorem},\footnote{%
Liouville theorem\emph{\ }(\ref{one}) is usually derived from the {%
continuity equation}
\begin{equation*}
\rho _{t}+\func{div}(\rho \,\mathbf{\dot{x}})=0\,.
\end{equation*}%
} written here in terms of Poisson brackets $\{.,.\}$:%
\begin{equation}
\rho _{t}=-\{\rho ,H\},  \label{one}
\end{equation}%
the phase-space density of the field theory associated with the matter DOF
of the MT--complex evolves with time as a consequence of phase-space
volume-preserving symmetries. More generally, statistical description of the
temporal evolution of the MT--complex using classical density matrices $\rho
(\phi ^{i},t)$ \cite{emn}:
\begin{equation}
\rho _{t}=-\{\rho ,H\}+\beta ^{i}G_{ij}\partial _{p_{j}}\rho \,,
\label{liouvmod}
\end{equation}%
where $p_{i}$ are momenta canonically-conjugate to the fields $\phi ^{i}$,
and $G_{ij}$ is the metric in the space of fields $\{\phi ^{i}\}$. The
non-Hamiltonian term in (\ref{liouvmod}) leads to a violation of the
Liouville theorem (\ref{one}) in the classical phase space $\{\phi
^{i},p_{j}\}$, and constitutes the basis for a dissipative
quantum-mechanical description of the system \cite{emn}, upon density-matrix
quantization. In string theory, summation over world sheet surfaces will
imply quantum fluctuations of the string target-space background fields $%
\phi ^{i}.$

Using Dirac's quantization rule: \ $\{{.,.}\}\longrightarrow -\mathrm{i}\,[\,%
{.,.}],$ the quantum version of (\ref{liouvmod}) reads (in terms of the
quantum commutator $[{.,.}])$ \cite{emn}:
\begin{equation}
{\hat{\rho}}_{t}=\mathrm{i}[{\hat{\rho}},{\hat{H}}]+\mathrm{i}\beta
^{i}G_{ij}[{\hat{\phi}}^{i},{\hat{\rho}}],  \label{liouQu}
\end{equation}%
where the hat denotes quantum operators, and appropriate \emph{quantum
ordering} (in the sense of \cite{lin}) is understood. In (\ref{liouQu}), $%
\hat{H}$ is the Hamiltonian evolution operator, while
\begin{equation*}
\hat{\rho}=\sum_{a}P(a)\left\vert \Psi _{a}\right\rangle \left\langle \Psi
_{a}\right\vert ,\quad \text{\textrm{with}}\quad (\mathrm{Tr}(\hat{\rho})=1),
\end{equation*}%
is von Neumann's \emph{density matrix} operator, in which each quantum state
$\left\vert \Psi _{a}\right\rangle $ occurs with probability $P(a)$; ~von
Neumann's entropy is defined as:
\begin{equation*}
S=-\mathrm{Tr}(\hat{\rho}[\ln \hat{\rho}]).
\end{equation*}

The very structure of the quantum Liouville equation (\ref{liouQu}) implies
the following properties of the MT--complex: \cite{emn,emn00,Nuovo94}:%
\newline
(i) conservation of probability $P$: \
\begin{equation*}
P_{t}=\mathrm{Tr}_{t}(\hat{\rho})=0;
\end{equation*}
(ii) conservation of average energy $\left\langle E\right\rangle $:\footnote{%
However, the quantum energy fluctuations:
\begin{equation*}
\delta E~\equiv ~[<<\mathcal{H}^{2}>>-(<<\mathcal{H}>>)^{2}]^{\frac{1}{2}}
\end{equation*}%
are time-dependent and actually decrease with time \cite{MN97}:
\begin{eqnarray*}
\partial _{t}(\delta E)^{2} &=&-i<<[\beta ^{i},\mathcal{H}]\beta
^{j}G_{ji}>>=<<\beta ^{j}G_{ji}\frac{d}{dt}\beta ^{i}>> \\
&=&-<<Q^{2}\beta ^{i}G_{ij}\beta ^{j}>>=-<<Q^{2}\partial _{t}C>>~\leq 0.
\end{eqnarray*}%
} \ $\left\langle E\right\rangle _{t}\equiv \mathrm{Tr}_{t}(\hat{\rho}%
E)=(p_{i}\beta ^{i})_{t}=0;$ and\newline
(iii) monotonic increase in entropy $S$:
\begin{equation*}
S_{t}\equiv -\mathrm{Tr}_{t}(\hat{\rho}\ln \hat{\rho})=S(\beta
^{i}G_{ij}\beta ^{j})\geq 0,
\end{equation*}%
-- which naturally implies a microscopic arrow of time within the
MT--complex \cite{Nuovo94}.

\subsection{SGE--solitons in neural impulse conduction}

Recently, two biophysicists from the Niels Bohr Institute in Copenhahen, T.
Heimburg and A. Jackson (see \cite{HJ1,HJ2}), challenged the half-a-century
old electrical theory of neural impulse conduction, proposed by A.L. Hodgkin
and A.F. Huxley in the form of their celebrated HH equations (1963 Nobel
Prize in Physiology or Medicine). The HH model, which relies on ionic
currents through ion channel proteins and the membrane capacitor, is the
presently accepted textbook model for the nerve impulse conduction.

For our readers' reference, here is a brief on the HH model, which (in its
basic form) consists of four coupled nonlinear first-order ODEs, including
the cable equation for the neural membrane potential $V$, together with $m,h$
and $n$ \ equations for the gating variables of Na and K channels and
leakage (see \cite{H-H,Hodgkin}; for recent reviews, see \cite{Guo12,TijHelen}):
\begin{eqnarray}
C_{m}\dot{V} &=&-g_{\mathrm{Na}}m^{3}h(V-V_{\mathrm{Na}})-g_{\mathrm{K}%
}n^{4}(V-V_{\mathrm{K}})-g_{\mathrm{L}}(V-V_{\mathrm{L}})+I_{j}^{\mathrm{ext}%
},  \notag \\
\dot{m} &=&-(a_{m}+b_{m})\,m+a_{m},\qquad \;\dot{h}=-(a_{h}+b_{h})\,h+a_{h},
\label{HH} \\
\dot{n} &=&-(a_{n}+b_{n})\,n+a_{n},\qquad \text{\textrm{where}}  \notag \\
a_{m} &=&0.1\,(V+40)/[1-\mathrm{e}^{-(V+40)/10}],\qquad \;b_{m}=4\,\mathrm{e}%
^{-(V+65)/18},  \notag \\
a_{n} &=&0.01\,(V+55)/[1-\mathrm{e}^{-(V+55)/10}],\qquad b_{n}=0.125\,%
\mathrm{e}^{-(V+65)/80},  \notag \\
a_{n} &=&0.07\,\mathrm{e}^{-(V+65)/20},\qquad \qquad b_{n}=1/[1+\mathrm{e}%
^{-(V+35)/10}].  \notag
\end{eqnarray}%
Here the reversal potentials of Na, K channels and leakage are: $V_{\mathrm{%
Na}}=50$ mV, $V_{\mathrm{K}}=-77$ mV and $V_{\mathrm{L}}=-54.5$ mV; the
maximum values of corresponding conductivities are: $g_{\mathrm{Na}}=120\;%
\mathrm{mS/cm}^{2}$, $g_{\mathrm{K}}=36\;\mathrm{mS/cm}^{2}$ and $g_{\mathrm{%
L}}=0.3\;\mathrm{mS/cm}^{2}$; the capacity of the membrane is: $C_{m}=1\;\mu
\mathrm{F/cm}^{2}$. The external, input current is given by:
\begin{equation}
I_{j}^{\mathrm{ext}}=g_{\mathrm{syn}}(V_{a}-V_{c})\sum_{n}\alpha (t-t_{in}),
\label{6}
\end{equation}%
which is induced by the pre-synaptic spike-train input applied to the neuron
$i$, given by: \ $U_{i}(t)=V_{a}\sum_{n}\delta (t-t_{in}).$ In (\ref{6}), $%
t_{in}$ is the $n$th firing time of the spike-train inputs, $g_{\mathrm{syn}%
} $ and $V_{c}$ denote the conductance and the reversal potential,
respectively, of the synapse, $\tau _{\mathrm{s}}$ is the time constant
relevant to the synapse conduction, and $\alpha (t)=(t/\tau _{\mathrm{s}})\;%
\mathrm{e}^{-t/\tau _{\mathrm{s}}}\Theta (t)\ ($where $\Theta (t)$ is the
Heaviside function).

In addition, Hodgkin and Huxley assumed that the total current is the sum of
the trans-membrane current and the current along the axon and that a
propagating solution exists that fulfills a wave equation, so the simple
cable ODE figuring in the basic HH model (\ref{HH}) was further expanded
into the following PDE for the propagating nerve impulse depending on the
axon radius $a$:
\begin{equation*}
\frac{a}{2R_{i}}V_{xx}=C_{m}V_{t}+g_{\mathrm{K}}(V-E_{K})+g_{\mathrm{Na}%
}(V-E_{Na}),
\end{equation*}%
where $R_{i}$ is the resistance of the cytosol within the nerve (see \cite%
{HJ2} for technical review).

The HH model was originally proposed to account for the property of squid
giant axons \cite{H-H,Hodgkin} and it has been generalized with
modifications of ion conductances. More generally, the HH--type models have
been widely adopted for a study on activities of \textit{transducer neurons}
such as motor and thalamus relay neurons, which transform the
amplitude-modulated input to spike-train outputs. For attempts to relate the
HH model (as well as its simplified form, Fitzhugh--Nagumo model (FHN)%
\footnote{%
Among several forms of the FHN--model, the simplest one (similar to the Van
der Pol oscillator) is suggested by FitzHugh \cite{FitzHugh}:
\begin{equation*}
\begin{array}{rcrcl}
\epsilon \frac{dx}{d\tau } & = & \epsilon \dot{x} & = & x-x^{3}-y, \\
\frac{dy}{d\tau } & = & \dot{y} & = & \gamma x-y+b,%
\end{array}%
\end{equation*}%
where $x$ is voltage (the fast variable), $y$ is the slow recovery variable
and $\gamma $, $b$, $\epsilon $ ($0\leq \epsilon \ll 1$) are parameters.}
\cite{FitzHugh,Nagumo}) with propagation of solitons in neural cell
membranes see \cite{Das95} and references therein.

However, the HH model fails to explain a number of features of the
propagating nerve pulse, including the reversible release and reabsorption
of heat\footnote{%
Electrical currents through resistors generate heat, independent of the
direction of the ion flux. The heat production in the HH--model should
always be positive, while the heat dissipation should be related to the
power of a circuit through the resistor, i.e. $\dot{Q}=P=V\cdot I>0$ (for
each of the conducting objects in all phases of the action potential; see
\cite{HJ1,HJ2}).} and the accompanying mechanical, fluorescence, and
turbidity changes \cite{HJ1,HJ2}:
\begin{quotation}
\textquotedblleft The most striking feature of the isothermal and isentropic
compression modulus is its significant undershoot and striking recovery.
These features lead generically to the conclusions (i) that there is a
minimum velocity of a soliton and (ii) that the soliton profiles are
remarkably stable as a function of the soliton velocity. There is a maximum
amplitude and a minimum velocity of the solitons that is close to the
propagation velocity in myelinated nerves..."
\end{quotation}

Earlier work of A.V. Hill \cite{Hill2} (another English Nobel laureate), on
heat production in nerves (which was based on his previous work on heat
production in contracted muscles \cite{Hill1}) is actually reviewed in \cite%
{Hodgkin}, where it is noted that the heat release and absorption response
during the action potential is important `but is not understood'.

Based on thermodynamic relation between heat capacity and membrane area
compressibility, Heimburg and Jackson considered in \cite{HJ1,HJ2} a (1+1)
hydrodynamic PDE for the dispersive sound propagation in a cylindrical
membrane of a density-pulse, governing the changes $\Delta \rho ^{A}$ (along
the $x$-axis) of the lateral membrane density $\rho ^{A}$, defined by: $%
\Delta \rho ^{A}(x,t)=\rho ^{A}(x,t)-\rho _{0}^{A}$\thinspace , where $\rho
_{0}^{A}=4.035\cdot 10^{-3}\,g/m^{2}$ is the equilibrium lateral area
density in the fluid phase of the membrane slightly above the melting point.
The related sound velocity $c$ can be expanded into a power series (close to
the lipid melting transition) as:
\begin{equation}
c^{2}=c_{0}^{2}+p(\Delta \rho ^{A})+q(\Delta \rho ^{A})^{2}+\ldots ,
\label{c2}
\end{equation}%
where $c_{0}=176.6$ m/s is the velocity of small amplitude sound, while $p$
and $q$ are parameters ($p=-16.6\,c_{0}^{2}/\rho _{0}^{A}$ and $%
q=79.5\,c_{0}^{2}/(\rho _{0}^{A})^{2}$).

In our standard $\phi -$notation, with $\phi (x,t)\equiv \Delta \rho
^{A}(x,t)$, the \emph{dispersive wave equation} of \cite{HJ1,HJ2}\ can be
written as:
\begin{equation}
\phi _{tt}=c^{2}\phi _{xx}-f(\phi )\,.  \label{waveHJ}
\end{equation}%
Here, we need to make two remarks regarding the dispersive wave equation (%
\ref{waveHJ}):
\begin{enumerate}
\item If the compressibility is approximately constant and if $\Delta \rho
^{A}\ll \rho _{0}^{A},$ then the dispersive force $f(\phi )$ is zero and (%
\ref{waveHJ}) reduces to the standard wave equation (depending only on the
small amplitude sound $c_{0}^{2}$): $\ $%
\begin{equation*}
\phi _{tt}=c_{0}^{2}\phi _{xx}\,.
\end{equation*}

\item If higher sound frequencies (resulting in higher propagation
velocities as the isentropic compressibility is a decreasing function of
frequency) become dominant, the dispersive forcing function $f(\phi )$ in (%
\ref{waveHJ}) needs to be defined, or \emph{ad-hoc chosen} \cite{HJ1,HJ2} to
mimic the linear frequency--dependence of the sound velocity with a positive
parameter $h$ as: \ $f(\phi )=h\phi _{xxxx\,}.$ In this case, the expansion (%
\ref{c2}) needs to be explicitely included into PDE (\ref{waveHJ}),
resulting in the equation governing dispersive sound propagation, which
reads (in original notation of \cite{HJ1,HJ2}):%
\begin{equation}
\frac{\partial ^{2}}{\partial t^{2}}\Delta \rho ^{A}=\frac{\partial }{%
\partial x}\left[ \left( c_{0}^{2}+p(\Delta \rho ^{A})+q(\Delta \rho
^{A})^{2}\right) \frac{\partial }{\partial x}\Delta \rho ^{A}\right] -h\frac{%
\partial ^{4}}{\partial x^{4}}\Delta \rho ^{A}.  \label{dispers}
\end{equation}%
Furthermore, by introducing the sound propagation velocity $v,$ after the
coordinate transformation: $z=x-v\cdot t$, the dispersive PDE (\ref{dispers}%
) can be recast into a time-independent form, describing the \emph{shape} of
a propagating density excitation:
\begin{equation*}
v^{2}\frac{\partial ^{2}}{\partial z^{2}}\Delta \rho ^{A}=\frac{\partial }{%
\partial z}\left[ \left( c_{0}^{2}+p(\Delta \rho ^{A})+q(\Delta \rho
^{A})^{2}\right) \frac{\partial }{\partial z}\Delta \rho ^{A}\right] -h\frac{%
\partial ^{4}}{\partial z^{4}}\Delta \rho ^{A}.
\end{equation*}%
This 1-dimensional PDE\ has a localized (stationary) solution \cite%
{Lautrup2005}:
\begin{equation*}
\Delta \rho ^{A}(z)=\frac{p}{q}\cdot \frac{1-\left( \frac{v^{2}-v_{min}^{2}}{%
c_{0}^{2}-v_{min}^{2}}\right) }{1+\left( 1+2\sqrt{\frac{v^{2}-v_{min}^{2}}{%
c_{0}^{2}-v_{min}^{2}}}\cosh \left( \frac{c_{0}}{h}z\sqrt{1-\frac{v^{2}}{%
c_{0}^{2}}}\right) \right) ,}
\end{equation*}%
which is a \emph{sech-type soliton,} a typical solution for KdV and NLS
equations. \qquad \qquad
\end{enumerate}

Now, without arguing either pro- or contra- Heimburg--Jackson theory of
\emph{neural sound propagation,} as an alternative to Hodgkin--Huxley \emph{%
electrical theory,} we will simply accept the natural solitary explanation
of the nerve impulse conduction, regardless of the physical medium that is
carrying it (sound, or heat, or electrical, or smectic liquid crystal \cite%
{Das95}, or possibly quantum-mechanical \cite{MN97}). However, we are free
to chose a different form for the dispersive force term $f(\phi )$ in the
perturbed wave equation (\ref{waveHJ}). For example, instead of the
Heimburg--Jackson's ad-hoc choice of the forth-derivative term, following
\cite{Sataric93} and subsequent studies of neural MTs, we can choose a
double-well quartic dispersive potential (\ref{dbwell}), which will,
combined with the approximation (\ref{sinTaylor}), result in the standard
SGE (\ref{SGE1}), generating analytical solutions of the traveling soliton,
kink-antikink and breather form (as described in the subsection \ref{kinks}
before).

\subsection{Muscular--contraction solitons on Poisson manifolds}
\label{Poisson}

For geometrical analysis of nonlinear PDEs, instead of using common
symplectic structures arising in ordinary Hamiltonian mechanics, the more
appropriate approach is a \textit{Poisson manifold} $(\mathbf{g}^{\ast
},\,\{F,G\})$, in which $\mathbf{g}^{\ast }$ is a chosen Lie algebra with a $%
(\pm )$ \textit{Lie--Poisson bracket} $\{F,G\}_{\pm }(\mu )$) and carries an
abstract \textit{Poisson evolution equation} $\dot{F}\,=\,\{F,H\}$. This
approach is well--defined in both the finite-- and the infinite--dimensional
case (see \cite{Marsden,GaneshSprBig,GaneshADG,GaneshControl}).

Recall that a \textit{Lie algebra} consists of a vector (e.g., Banach) space
$\mathbf{g}$ carrying a bilinear skew--symmetric operation $[,]:\mathbf{{g}%
\times {g}\rightarrow {g}}$, called the \textit{commutator} or Lie bracket.
This represents a pairing:\footnote{%
Let $E_{1}$ and $E_{2}$ be Banach spaces. A continuous bilinear functional $%
<,>:E_{1}\times E_{2}\rightarrow \mathbb{R}$ is nondegenerate if $%
<x,y>\,=\,0 $ implies $x\,=\,0$ and $y\,=\,0$ for all $x\in E_{1}$ and $y\in
E_{2}$. We say $E_{1}$ and $E_{2}$ are in \emph{duality} if there is a
nondegenerate bilinear functional $<,>:E_{1}\times E_{2}\rightarrow \mathbb{R%
}$. This functional is also referred to as an $L^{2}-$\emph{pairing} of $%
E_{1}$ with $E_{2}$.} $[\xi ,\eta ]\,=\,\xi \eta \,-\,\eta \xi $ \ of
elements $\xi ,\eta \in \mathbf{g}$ and satisfies \textit{Jacobi identity:}
\begin{equation*}
\lbrack \lbrack \xi ,\eta ],\mu ]+[[\eta ,\mu ],\xi ]+[[\mu ,\xi ],\eta ]=0.
\end{equation*}

Let $\mathbf{g}$ be a (finite-- or infinite--dimensional) Lie algebra and $%
\mathbf{g}^{\ast }$ its dual Lie algebra, that is, the vector space $L^{2}$
paired with $\mathbf{g}$ \emph{via} the inner product $<,>:\mathbf{g}^{\ast
}\times \mathbf{g}\rightarrow \mathbb{R}$. If $\mathbf{g}$ is
finite--dimensional, this pairing reduces to the usual action (interior
product) of forms on vectors. The standard way of describing any
finite--dimensional Lie algebra $\mathbf{g}$ is to provide its $n^{3}$ \emph{%
structural constants} $\gamma _{ij}^{k}$, defined by $[\xi _{i},\xi
_{j}]\,=\,\gamma _{ij}^{k}\xi _{k}$, in some basis $\xi _{i},\,(i=1,\dots
,n) $

For any two smooth functions $F,G:\mathbf{g}^{\ast }\rightarrow \mathbb{R}$,
we define the ($\pm $) \emph{Lie--Poisson bracket} by:\footnote{%
The $(\pm )$ Lie--Poisson bracket (\ref{LPB}) is a bilinear and
skew--symmetric operation. It also satisfies the Jacobi identity:
\begin{equation*}
\{\{F,G\},H\}_{\pm }(\mu )+\{\{G,H\},F\}_{\pm }(\mu )+\{\{H,F\},G\}_{\pm
}(\mu )=0
\end{equation*}%
(thus confirming that $\mathbf{{g}^{\ast }}$ is a Lie algebra), as well as
the Leibniz rule:
\begin{equation}
\{FG,H\}_{\pm }(\mu )=F\{G,H\}_{\pm }(\mu )\,+\,G\{F,H\}_{\pm }(\mu ).
\label{PBm}
\end{equation}%
}
\begin{equation}
\{F,G\}_{\pm }(\mu )=\pm <\mu ,\left[ {\frac{{\delta F}}{{\delta \mu }}},{%
\frac{{\delta G}}{{\delta \mu }}}\right] >.  \label{LPB}
\end{equation}%
Here $\mu \in \mathbf{g}^{\ast }$, $[\xi ,\mu ]$ is the Lie bracket in $%
\mathbf{g}$ and ${{\delta F}/{\delta \mu }}$, ${{\delta G}/{\delta \mu }}\in
\mathbf{g}$ are the functional derivatives\footnote{%
For any two smooth functions $F:\mathbf{{g}^{\ast }\rightarrow \mathbb{R}}$,
we define the \textit{Fr\'{e}chet derivative} $D$ on the space $L(\mathbf{g}%
^{\ast },\mathbb{R})$ of all linear diffeomorphisms from $\mathbf{g}^{\ast }$
to $\mathbb{R}$ as a map $DF:\mathbf{g}^{\ast }\rightarrow L(\mathbf{g}%
^{\ast },\mathbb{R});\,\,\mu \mapsto DF(\mu )$. Further, we define the
\textit{functional derivative} ${{\delta F}/{\delta \mu }}\in \mathbf{g}$ by
\begin{equation*}
DF(\mu )\cdot \delta \mu \,=\,<\delta \mu ,{\frac{{\delta F}}{{\delta \mu }}}%
>
\end{equation*}%
with arbitrary `variations' $\delta \mu \in \mathbf{g}^{\ast }$.} of $F$ and
$G$.

Given a smooth Hamiltonian function $H:\mathbf{g^{\ast }}\rightarrow \mathbb{%
R}$ on the Poisson manifold $(\mathbf{g}^{\ast },\{F,G\}_{\pm }(\mu ))$, the
time evolution of any smooth function $F:\mathbf{g^{\ast }}\rightarrow
\mathbb{R}$ is given by the abstract \emph{Poisson evolution equation:}
\begin{equation}
\dot{F}=\{F,H\}.  \label{PoissonEq}
\end{equation}

Now, the basis of Davydov's molecular model of muscular contraction\footnote{%
For general overview of muscular contraction physiology and mechanics, see
\cite{HBE,NatBio}.} is oscillations of Amid I peptide groups with associated
dipole electric momentum inside a spiral structure of myosin filament
molecules (see \cite{Davydov73,DavydovBiol,DavydovSolit}). There is a
simultaneous resonant interaction and strain interaction generating a
collective interaction directed along the axis of the spiral. The resonance
excitation jumping from one peptide group to another can be represented as
an exciton, the local molecule strain caused by the static effect of
excitation as a phonon and the resultant collective interaction as a \textit{%
soliton}.

Davydov's own model of muscular solitons was given by the \textit{nonlinear
Schr\"{o}dinger equation} (NLS) \cite{Davydov73,DavydovBiol}:\footnote{%
For a different (financial) application, with a variety of traveling-wave
solutions of the NLS (\ref{NSchr}), including both sech- and tanh-solitons,
solved in terms of Jacobi elliptic functions, see\cite{ivopt10,ivopt11} and
references therein.}
\begin{equation}
\mathrm{i}\psi _{t}=-\psi _{xx}+2\chi |\psi |^{2}\psi ,  \label{NSchr}
\end{equation}%
for $-\infty <x<+\infty $. Here $\psi =\psi (x,t)$ is a smooth
complex-valued wave function with initial condition $\psi
(x,t)|_{t=0}\,=\,\psi (x)$ and $\chi $ is a nonlinear parameter. In the
linear limit ($\chi \,=\,0$) (\ref{NSchr}) becomes the ordinary Schr\"{o}%
dinger equation for the wave $\psi $-function of the free 1D particle with
mass $m=1/2$ \cite{DavydovBiol}.

To put this muscular--contraction model into a rigorous geometrical
settings, we can define the infinite-dimensional phase-space manifold $%
\mathcal{P}=\{(\psi ,\bar{\psi})\in S(\mathbb{R}\mathbf{,{C})\}}$, where $S(%
\mathbb{R}\mathbf{,{C})}$ is the Schwartz space of rapidly-decreasing
complex-valued functions defined on $\mathbb{R}$). We define also the
algebra $\mathbf{{\chi }(\mathcal{P})}$ of observables on $\mathcal{P}$
consisting of real-analytic functional derivatives ${\delta F}/{\delta \psi }
$, ${\delta F}/{\delta \bar{\psi}}\in S(\mathbb{R}\mathbf{,{C})}$. The
Hamiltonian function $H:\mathcal{P}\rightarrow \mathbb{R}$ is given by:
\begin{equation*}
H(\psi )\,=\,\int_{-\infty }^{+\infty }\left( \left\vert \psi
_{x}\right\vert ^{2}\,+\,\chi |\psi |^{4}\right) dx
\end{equation*}%
and is equal to the total energy of the soliton, which is a conserved
quantity for (\ref{NSchr}) (see, e.g. \cite{Seiler1,Seiler2}).

The Poisson bracket on $\mathbf{{\chi }(\mathcal{P})}$ represents a direct
generalization of the classical finite--dimensional Poisson bracket \cite%
{GaneshPoisson}:
\begin{equation}
\{F,G\}_{+}(\psi )\,=\,\mathrm{i}\int_{-\infty }^{+\infty }\left( \frac{%
\delta F}{\delta \psi }\frac{\delta G}{\delta \bar{\psi}}\,-\,\frac{\delta F%
}{\delta \bar{\psi}}\frac{\delta G}{\delta \psi }\right) dx.  \label{xP}
\end{equation}%
It manifestly exhibits skew--symmetry and satisfies Jacobi identity. The
functional derivatives are given by:
\begin{equation*}
{\delta F}/{\delta \psi }\,=\,-\mathrm{i}\{F,\bar{\psi}\}\qquad \mathrm{%
and\qquad }{\delta F}/{\delta \bar{\psi}}\,=\,\mathrm{i}\{F,\psi \}.
\end{equation*}%
Therefore the algebra of observables $\mathbf{{\chi }(\mathcal{P})}$
represents the Lie algebra and the Poisson bracket is the $(+)$ Lie--Poisson
bracket $\{F,G\}_{+}(\psi )$. The nonlinear Schr\"{o}dinger equation (\ref%
{NSchr}) for the solitary particle--wave is a Hamiltonian system on the Lie
algebra $\mathbf{{\chi }(\mathcal{P})}$ relative to the $(+)$ Lie--Poisson
bracket $\{F,G\}_{+}(\psi )$ and Hamiltonian function $H(\psi )$. Therefore
the Poisson manifold $(\mathbf{{\chi }(\mathcal{P})}$, $\{F,G\}_{+}(\psi ))$
is defined and the abstract Poisson evolution equation (\ref{PoissonEq}),
rewritten here as: $\ {\dot{\psi}}=\{{\psi },H\},$ which holds for any
smooth function ${\psi }:\mathbf{{\chi }(\mathcal{P})\rightarrow }\mathbb{R}$%
, is equivalent to (\ref{NSchr}).

An alternative model of muscular soliton dynamics is provided by the \textit{%
Korteweg--deVries equation} (KdV, see \cite{GaneshPoisson}):\footnote{%
The most common traveling-wave solutions of the KdV (\ref{KdV1}) are
sech-solitons with the velocity $c$, of the form \cite{Terng}:
\par
\begin{equation*}
f(x,t)=\frac{c}{2}\mathrm{sech}^{2}\left[ \frac{\sqrt{c}}{2}(x-ct)\right] .
\end{equation*}%
}
\begin{equation}
f_{t}\,-\,6ff_{x}\,+\,f_{xxx}\,=\,0,\qquad (f_{x}=\partial _{x}f),
\label{KdV1}
\end{equation}%
where $x\in \mathbb{R}$ and $f$ is a real--valued smooth function defined on
$\mathbb{R}$. This equation is related to the ordinary Schr{\"{o}}dinger
equation by the inverse scattering method (see \cite{Seiler1,Seiler2}).

Again, we may define the infinite--dimensional phase--space manifold $%
\mathcal{V}=\{f\in S(\mathbb{R})\}$, where $S(\mathbb{R}$) is the Schwartz
space of rapidly--decreasing real--valued functions $\mathbb{R}$). Further,
we define $\mathbf{{\chi }(\mathcal{V})}$ to be the algebra of observables
consisting of functional derivatives ${\delta F}/{\delta f}\in S(\mathbb{{R})%
}$. The Hamiltonian $H:\mathcal{V}\rightarrow \mathbb{R}$ is given by:
\begin{equation*}
H(f)\,=\,\int_{-\infty }^{+\infty }(f^{3}\,+\,\frac{1}{2}f_{x}^{2})\,dx,
\end{equation*}%
and provides the total energy of the soliton, which is a conserved quantity
for (\ref{KdV1}) (see \cite{Seiler1,Seiler2}).

As a real--valued analogue to (\ref{xP}), the $(+)$ Lie--Poisson bracket on $%
\mathbf{{\chi }(\mathcal{V})}$ is given \emph{via} (\ref{PBm}) by:
\begin{equation*}
\{F,G\}_{+}(f)\,=\,\int_{-\infty }^{+\infty }\frac{\delta F}{\delta f}\frac{d%
}{dx}\frac{\delta G}{\delta f}\,dx.
\end{equation*}%
Again it possesses skew--symmetry and satisfies Jacobi identity. The
functional derivatives are given by ${\delta F}/{\delta f}\,=\,\{F,f\}$. The
KdV equation (\ref{KdV1}), describing the behavior of the muscular molecular
soliton, is a Hamiltonian system on the Lie algebra $\mathbf{{\chi }(%
\mathcal{V})}$ relative to the $(+)$ Lie--Poisson bracket $\{F,G\}_{+}(f)$
and the Hamiltonian function $H(f)$. Therefore, the Poisson manifold $(%
\mathbf{\chi }(\mathcal{V}),\{F,G\}_{+}(f))$ is defined and the abstract
Poisson evolution equation (\ref{PoissonEq}), rewritten here as: \ $\dot{f}%
=\{f,H\},$ which holds for any smooth function $f:\mathbf{{\chi }(\mathcal{V}%
)\rightarrow }\mathbb{R}$, is equivalent to (\ref{KdV1}).

Another alternative model of muscular soliton dynamics is provided by our
SGE (\ref{SGE1}): $\ \phi _{tt}=\phi _{xx}-\sin \phi $. Again, we may define
the infinite--dimensional phase--space manifold $\mathcal{V}=\{\phi \in S(%
\mathbb{R})\}$, where $S(\mathbb{R}$) is the Schwartz space of
rapidly--decreasing real--valued functions $\mathbb{R}$). Further, we define
$\mathbf{{\chi }(\mathcal{V})}$ to be the algebra of observables consisting
of functional derivatives ${\delta F}/{\delta }\phi \in S(\mathbb{{R})}$.
The Hamiltonian $H:\mathcal{V}\rightarrow \mathbb{R}$ is given by:%
\begin{equation*}
H(\phi )=\int_{-\infty }^{\infty }\left[ \frac{1}{2}(\pi ^{2}+\phi
_{x}^{2})+1-\cos \phi \right] \,dx,
\end{equation*}%
and provides the total energy of the soliton, which is a conserved quantity
for the SGE (\ref{SGE1}). The $(+)$ Lie--Poisson bracket on $\mathbf{{\chi }(%
\mathcal{V})}$ is given \emph{via} (\ref{PBm}) by:
\begin{equation*}
\{F,G\}_{+}(\phi )\,=\,\int_{-\infty }^{\infty }\left( \frac{\delta F}{%
\delta \phi }\frac{\delta G}{\delta \pi }-\frac{\delta F}{\delta \pi }\frac{%
\delta G}{\delta \phi }\right) dx.
\end{equation*}%
Again it possesses skew--symmetry and satisfies Jacobi identity. The
functional derivatives are given by: ${\delta F}/{\delta }\phi =\,\{F,\phi
\}\in S(\mathbb{{R})}$. The SGE (\ref{SGE1}), describing the behavior of the
molecular muscular soliton, is a Hamiltonian system on the Lie algebra $%
\mathbf{{\chi }(\mathcal{V})}$ relative to the $(+)$ Lie--Poisson bracket $%
\{F,G\}_{+}(\phi )$ and the Hamiltonian function $H(\phi )$. Therefore, the
Poisson manifold $(\mathbf{\chi }(\mathcal{V}),\{F,G\}_{+}(\phi ))$ is
defined and the abstract Poisson evolution equation (\ref{PoissonEq}),
rewritten here as: \ $\dot{\phi}=\{\phi ,H\},$ which holds for any smooth
function $\phi :\mathbf{{\chi }(\mathcal{V})\rightarrow }\mathbb{R}$, is
equivalent to (\ref{SGE1}).

\section{Conclusion}

In this paper, we have reviewed sine--Gordon equation and its traveling wave solutions. These solitary spatiotemporal processes can serve as realistic models of nonlinear excitations in complex systems in physical sciences as well as in various living cellular structures, including intra--cellular ones (DNA, protein folding and microtubules) and inter--cellular ones (neural impulses and muscular contractions). We have showed that sine--Gordon solitons, kinks and breathers can give us new insights even in such long--time established and Nobel--Prize winning living systems as the Watson--Crick double helix DNA model and the Hodgkin--Huxley neural conduction model.\\

\bigbreak

\end{document}